\documentclass[prb,aps,amsfonts,amssymb,floatfix,showpacs,superscriptaddress]{revtex4}
\usepackage{graphicx}
\usepackage{dcolumn}
\usepackage{bm}
\usepackage{color} 
\usepackage{epsfig}
\usepackage{epstopdf}
\begin{document} 
\title{
Entropic Origin of Pseudogap Physics and a Mott-Slater Transition in Cuprates}
\author{R.S. Markiewicz{\footnote{Correspondence to markiewic@neu.edu}}, I.G. Buda, P. Mistark, C. Lane and A. Bansil}
\affiliation{ Physics Department, Northeastern University, Boston MA 02115, USA}
\begin{abstract}
We propose a new approach to understand the origin of the pseudogap in the cuprates, in terms of bosonic entropy.  The near-simultaneous softening of a large number of different $q$-bosons yields an extended range of short-range order, wherein the growth of magnetic correlations with decreasing temperature $T$ is anomalously slow.  These entropic effects cause the spectral weight associated with the Van Hove singularity (VHS) to shift rapidly and nearly linearly toward half filling at higher $T$, consistent with a picture of the VHS driving the pseudogap transition at a temperature $\sim T^*$.  As a byproduct, we develop an order-parameter classification scheme that predicts supertransitions between families of order parameters.  As one example, we find that by tuning the hopping parameters, it is possible to drive the cuprates across a {\it transition between Mott and Slater physics}, where a spin-frustrated state emerges at the crossover.  
\end{abstract} 
\maketitle

\section*{Introduction}

Evidence is growing that the `pseudogap phase' found in cuprates is home to one or more competing phases, including a variety of stripe, spin-, or charge-density wave (S/CDW) phases.\cite{1Kive,2Vojt,3Wu,4Ghir,5Achk,6Chan,7LeBo,8Blac,9Doir,10Comi,11daSi,12Fuji}  The CDW phase, in particular, has stimulated considerable interest\cite{13Metl,14Wang,15Efet,16Meie,17LaPl,18Hayw,19Bulu,20Alla,21Fuji,22RSMch}. Many of these phases, including superconductivity, seem to appear at temperatures well below the pseudogap temperature $T^*$, so the exact relation between the pseudogap and these other phases remains elusive.   Indeed, the real puzzle is understanding why the pseudogap phase bears so little resemblance to a conventional phase transition.  Here we demonstrate that the pseudogap phenomenon arises from strong mode coupling, where a large number of density waves with similar ordering wave-vectors $q$ attempt to soften and condense at the same time.  The associated entropy leads to anomalously low transition temperatures and extended ranges of short-range order, characteristic of the pseudogap phase.  In an extreme case, this can lead to a {\it condensation bottleneck} where long-range order is absent,  leaving behind a novel entangled spin-frustrated phase.  

To understand how entropy drives short range order, consider the Hubbard model with on-site repulsion $U$, nearest neighbor hopping energy $t$, and exchange $J=4t^2/U$.   A simple mean field solution correctly captures an antiferromagnetic ground state with ordering vector $q=(\pi,\pi)$ and gap $2\Delta \sim U$.   However, the same calculation yields a Neel temperature $T_N\sim U$, instead of the strong coupling result, $\sim 1/U$.  The resolution of this problem is the electronic entropy:  
repeating the calculation for magnetic order with any other $q$-vector, including ferromagnetism ($q=0$) also produces a gap of order $U$.  While the $(\pi,\pi)$ state is the ground state, the other states are higher in energy by no more than a factor $\sim J$.  Hence, at low $T$ the energy wins out, and the system condenses into its ground state, while at higher temperature entropy wins and the system lowers its free energy by maximizing its entropy, forming a mixture of all these competing phases.  Thus, longe-range order is lost at a $T_N\sim J$.\cite{23RSMMC}  The problem is how to incorporate this entropic effect into a calculation of the phase diagram.   Similar problems have arisen in the past.  In Overhauser's theory of CDWs in alkali metals,\cite{24Over} the $q$-vectors for all points on a spherical Fermi surface become unstable simultaneously and cannot be handled one at a time.  These effects are often referred to as phonon entropy \cite{25McMi}, or more generally as boson entropy, and can be analyzed via vertex corrections which take proper account of mode coupling\cite{26Moti,27Yosh}.   Here we develop a similar theory for the cuprates via a self-consistent renormalization calculation of the vertex corrections.  Similar effects are likely to be present in many other families of correlated materials. 

The ultimate origin of the bosonic entropy is localization. We provide two examples of this.  First, in the above discussion of AFM,  large $U$ means single spins tend to localize on individual copper atoms and the effective hopping is renormalized from $t$- to $J$-scales, leading to much flatter dispersions, and to small differences in energy between different configurations of spins. For our second example, we consider a situation where an electron-hole pair --our electronic boson--can form a bound state, an exciton. When excitons are strongly bound, they can condense, creating a form of a CDW or SDW \cite{29Bron,30Cote,28Halp}.  In the Bose condensation of excitons, the mean-field transition is found to correspond to the temperature at which excitons are formed.  However, since the excitons are localized in real space, they are greatly spread in $q$.  When fluctuations are included, the real transition lies at a much lower $T$, when all excitons condense into the lowest $q$ state. 

This paper is organized in the following Sections:  Key findings, Methods, Results (DFT-Lindhard), Results (Beyond RPA), and Discussion.  The Results section has been split into two parts.  The first [DFT-Lindhard susceptibility] shows that the Lindhard susceptibility can be separated into two components, one fermionic, associated with Fermi surface nesting, and more important at low $T$, the other bosonic, associated with Van Hove singularity (VHS) nesting, and usually more important at higher $T$, where Pauli blocking becomes unimportant.  The second Results section [Beyond RPA susceptibility] shows how the competition between these components can be incorporated into Moriya-Hertz-Millis theory by extending the calculations to realistic band structures.
A number of specialized issues are covered in the Supplementary Material.

\begin{figure}
\leavevmode
\rotatebox{0}{\scalebox{0.84}{\includegraphics{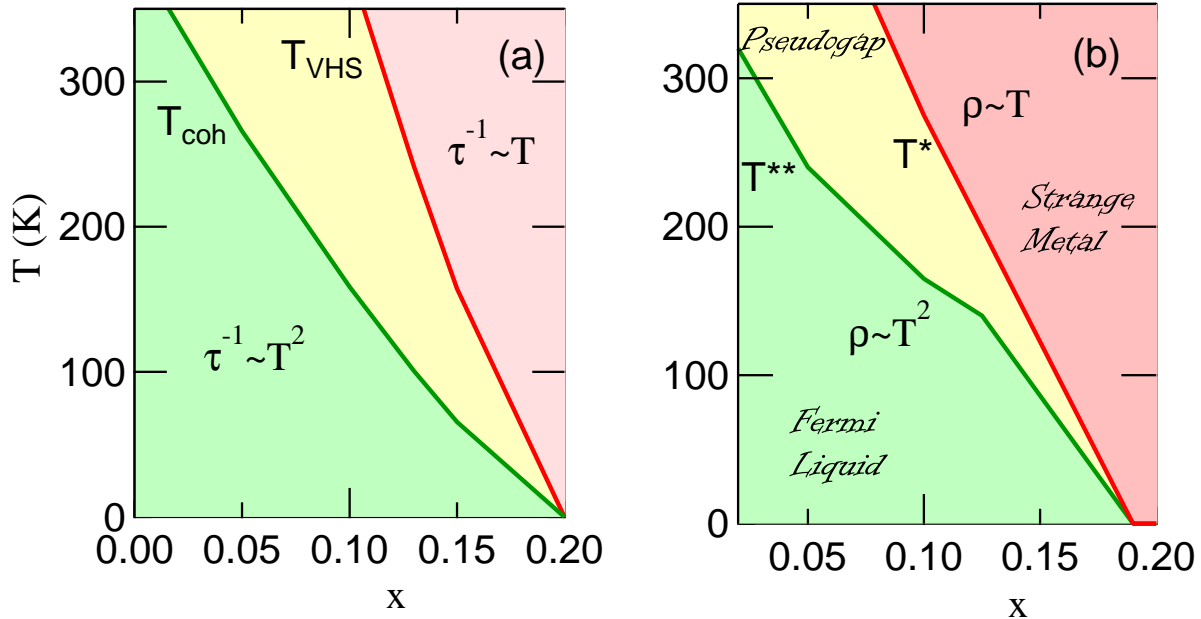}}}
\vskip0.5cm  
\caption{
{\bf 
Phase diagram of LSCO, comparing VHS-phase diagram (a) with experimental pseudogap phase diagram\cite{9Doir} (b).}
}
\label{fig:2cN}
\end{figure}

\section*{Key findings}

\subsection*{DFT-Lindhard susceptibility}

Our study illustrates the transformative role of the density functional theory (DFT) in many-body perturbation theory (MBPT)\cite{61Hedi}, and provides a deeper understanding of the bare (Lindhard) susceptibility $\chi_0$ [Methods Section]. Textbook calculations of $\chi_0$ replace all band structure effects by a simple electron gas with a parabolic band.  When realistic band structures are included, the Lindhard function becomes a nesting map\cite{Gzm1}, showing all the weighted nesting vectors of the various Fermi surfaces -- in particular, a scaled map of each Fermi surface, $q=2k_F$ (where $k_F$ is the Fermi wave vector), associated with intraband nesting [Results Section (DFT-Lindhard)].  Indeed, we find that $\chi_0$ can serve as a basis for classifying phase transitions, analogous to the spectrum generating algebras of nuclear physics [Supplemental Materials Section I].  We find that doping-temperature phase space can be divided up into fluctuation maps, or {\it basins of attraction}, where the character of the dominant nesting vector changes. These fluctuation maps lead to the idea of {\it equivalence classes}, where materials with very different sets of hopping parameters nevertheless display topologically equivalent fluctuation maps.  In particular, we find that LSCO lies in a different equivalence class from other cuprates.  From each equivalence class, we can define a {\it reference family} having a minimal number of hopping parameters, that can be used to interpolate between different cuprates.  In particular, we find that all cuprates lie on a single line in hopping parameter space, and that this tuning allows us to understand that LSCO is more strongly correlated than other cuprates. However, the susceptibility differs in one key way from most other properties of a Fermi liquid.  While many properties are controlled entirely by states near the Fermi level, the susceptibility can also contain important contributions from states deep in the Fermi sea, here referred to as bulk contributions.  In the cuprates, we find that $\chi_0$ contains not only sharp features associated with the Fermi surface nesting map, but also contains a smoothly varying background, peaking at $q=(\pi,\pi)$ and giving rise to the near-$(\pi,\pi)$-plateau in the susceptibility.  This background is associated with the  VHS peak, which lies a distance $T_{VHS}=(E_F-E_{VHS})/k_B$ below the Fermi level, Fig.~\ref{fig:2cN}(a).  Since it peaks away from the Fermi level, the corresponding features have a strong temperature dependence, associated with the smearing of the Fermi functions that enter in the expression for $\chi_0$, an effect referred to as Pauli blocking.

\subsection*{Role of entropy and VHS nesting in pseudogap physics}

McMillan's phononic entropy referred to a competition wherein phonons with many different $q$-vectors try to soften at the same time\cite{25McMi}, preventing any one phonon from condensing until very low $T$ and giving rise to an extended range of short-range order.  We find that similar strong entropic effects arise from electronic bosons (electron-hole pairs), predominantly associated with the VHS.  In turn, the entropic effects modify VHS physics, causing the main spectral weight associated with the VHS to shift rapidly towards half filling as $T$ increases, not following the evolution of the density of states (DOS).  This anomalous shift suggests that the VHS may play a large role in pseudogap physics, Fig.~\ref{fig:2cN}. \cite{9Doir} The key anomaly in pseudogap physics lies not below the pseudogap temperature but above $T^*$,  Fig.~\ref{fig:2cN}(b), where the resistivity in the strange metal phase varies linearly with $T$, $\rho\sim T$.  The pseudogap temperature $T^*$ indicates the onset of deviations from linear resistivity, while below a lower temperature $T^{**}$ the resistivity becomes purely quadratic, suggestive of a Fermi liquid regime.  While the strange metal phase is suggestive of the quantum critical phase in the Heisenberg quantum phase transition\cite{HCN}, there is no obvious reason why the resistivity of the Heisenberg model should be linear in $T$, nor any natural candidate for either the pseudogap (since the Heisenberg model has a well-defined ordered phase), or for the transition at $T^{**}$. In contrast, these features arise naturally in our VHS model, Fig.~\ref{fig:2cN}(a).  For $T>T_{VHS}$ Pauli blocking is completely ineffective, leaving the expected VHS scattering rate $\tau^{-1}\sim T$.  Below $T_{VHS}$ this blocking of the VHS starts to turn on, while coherent Fermi surface features appear, until below a coherent-incoherent crossover, $T<T_{coh}$, the VHS scattering is suppressed, leaving behind a Fermi liquid $\tau^{-1}\sim T^2$.   These results are consistent with experimental findings that the pseudogaps in Bi2201 and Bi2212 terminate near the VHS doping $x_{VHS}$\cite{47Piri,48Niem,49Benh}.

\begin{figure}
\leavevmode
\rotatebox{0}{\scalebox{0.54}{\includegraphics{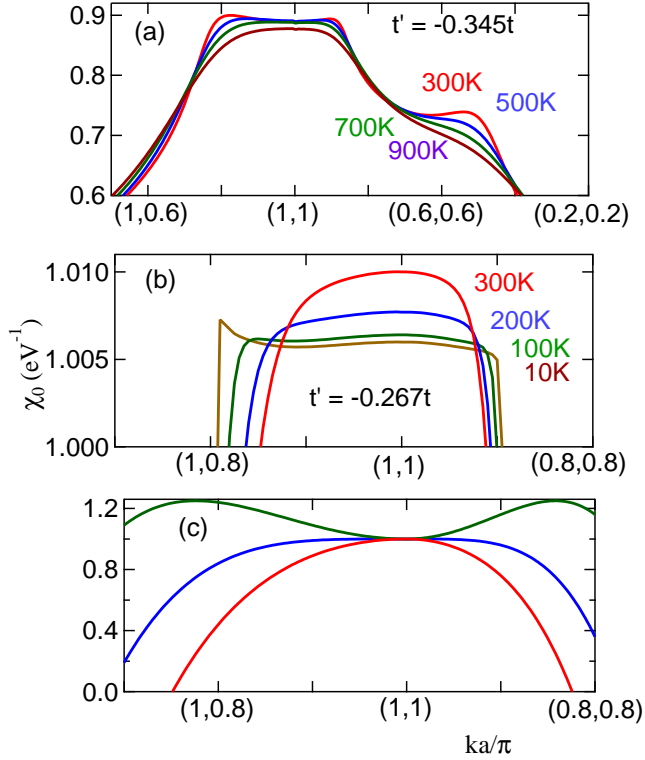}}}
\vskip0.5cm    
\caption{
{\bf Susceptibility near commensurate-incommensurate transition in $t-t'$-model,} for (a) $t'=-0.345t$ and (b) $t'=-0.267t$.  (c) A simple `Mexican hat' model, $\chi_=1-(A_2\bar q^2+A_4\bar q^4)$, for $\bar q =(\pi,\pi)-q$, with $A_4>0$ and varying $A_2$: $A_2>0$ (red line), =0 (blue line), and $<0$ (green line). 
} 
\label{fig:3g2N}
\end{figure}

\subsection*{Mode coupling and order parameter competition}

The bosonic entropy can be incorporated into the DFT-Lindhard function by modifying  Moriya's mode-coupling approach\cite{39Naga}to include realistic band-structures via a susceptibility DOS [Results Section (Beyond RPA)].  We use this technique to calculate the growth of the correlation length in LSCO, finding that the Mermin-Wagner theorem is satisfied leading to a broad regime of short-range order. We also find a temperature range where the correlation length does not grow at al. By tuning the hopping parameters, we find that this is associated with a  `supertransition' between two families of order parameters.  This is particularly interesting as it involves competition between $(\pi,\pi)$ Mott antiferromagnetic order driven by VHS fluctuations and a $(\pi,\pi-\delta)$ Slater SDW driven by conventional Fermi surface nesting.  We are able to define an {\it order parameter DOS} that diverges at the transition: Fig.~\ref{fig:3g2N} shows the susceptibility near the transition for two different values of $t'/t$.  At the transition, the susceptibility at $(\pi,\pi)$ changes from a maximum to a local minimum, and the susceptibility is quite flat across the entire $(\pi,\pi)$-plateau.  This is especially true in frame (b), which is tuned close to the $T=0$ transition (note change of scale).  It is convenient to introduce a susceptibility density-of-states (DOS) to quantify the number of competing states.  This behavior can be approximately captured in a simple `Mexican hat' model,  Fig.~\ref{fig:3g2N}(c).  A similar analysis for Bi2201 finds a transition between $(\pi,\pi-\delta)$ and antinodal nesting order, consistent with the SDW-to-CDW transition found in most cuprates.

\section*{Methods}

Our calculation is a form of many-body perturbation theory (MBPT) based on Hedin's scheme\cite{61Hedi}.  The scheme involves four elements: electrons are described by Green's functions $G$ with DFT-based dispersions renormalized by a self-energy $\Sigma$; electronic bosons [electron-hole pairs] are described by a spectral weight [susceptibility] $\chi$ renormalized by vertex corrections $\Gamma$.  Neglecting vertex corrections, the self-energy can be calculated as a convolution of $G$ and $W=U^2\chi$ (GW approximation).  This approach has been used to solve the energy gap problem in semiconductors, where the $\Gamma$ correction leads to excitons via the solution of a Bethe-Salpeter equation\cite{62Rohl,63Onid}, and in extending DMFT calculations to incorporate more correlations [e.g., DMFT+GW, etc.].  Our approach here is to extend our previous GW calculations [quasiparticle-GW or QPGW\cite{AIP}] to approximately include vertex corrections.  

In QPGW we introduce an auxiliary function $G_Z=Z/(\omega-\epsilon_{\bf k}^{QP})$, where the dressed, or QP dispersion is $\epsilon_{\bf k}^{QP}=Z\epsilon_{\bf k}^{DFT}$, and $\epsilon_{\bf k}^{DFT}$ is the bare, or DFT dispersion. $G_Z$ behaves like the Green's function of a Landau-type QP -- a free electron with renormalized parameters that describes the low-energy dressed electronic excitations.  However, this is a non-Fermi liquid type QP, since the frequency-integral of $Im(G_Z)$ is $Z$ and not unity.  That is, the $Z$-QP describes only the coherent part of the electronic dispersion, and is not in a 1:1 correspondence with the original electrons.  The importance of such a correction can be readily demonstrated.  Since a Z-QP has only the weight $Z$ of a regular electron, the susceptibility [a convolution of two $G$s] is weaker by a factor of $Z^2$ than an ordinary bare susceptibility.  To match this effect in the Stoner criterion requires introducing an effective $U_{eff}=ZU$.  In contrast, MBPT calculations in semiconductors typically set the GW-corrected Green's function to $G_{GW}^{-1}=\omega-\epsilon_{\bf k}^{QP}$, where $\epsilon_{\bf k}^{QP}$ is the average GW-renormalized dispersion\cite{64Bech}, thereby missing the reduced spectral weight of the low-energy, coherent part of the band.

In this paper the DFT-Lindhard susceptibility $\chi_0$ is calculated at intermediate coupling level, using $Z$-corrected first-principles dispersions to describe the coherent part of the quasiparticle-GW dressed carriers.\cite{AIP}  We calculate $\chi_0$ based on the dressed dispersion $\epsilon_Z=Z\epsilon_0$, assuming a doping-independent $Z=0.5$.  
We use a single-band model of the cuprates, working in a purely magnetic sector, with the Hubbard $U$ controlling all fluctuations; there is a competition in this model between near-nodal and antinodal nesting (ANN) which mimics the SDW-CDW competition in cuprates, sharing the same nesting vectors\cite{22RSMch,Gzm1}.

\section*{Results: DFT-Lindhard susceptibility}

\subsection*{Sorting the susceptibility peaks: basins of attraction for cuprates}
\subsubsection*{Fluctuation maps}

\begin{figure}
\leavevmode  
\rotatebox{0}{\scalebox{0.45}{\includegraphics{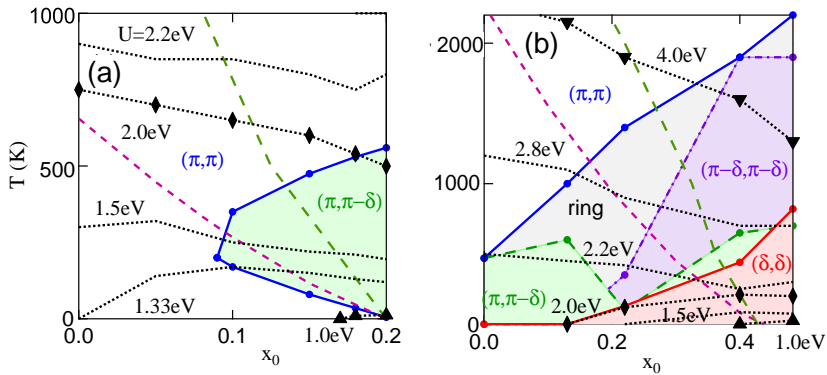}}}
\caption{
{\bf Mean-field phase diagrams of cuprates}. DFT-based models of LSCO (a) and Bi2201 (b).  Black dots indicate mean-field phase boundaries for different $U$ values, including $U$ = 4 (down triangles), 2 (diamonds), or 1~eV (up triangles).  Here $x_0$ is the doping at $T=0$ (plots are at constant $E_F$), but for qualitative purposes we can assume $x\simeq x_0$.  
} 
\label{fig:009c} 
\end{figure}
We define fluctuation maps in x-T space as maps of domains (`basins of attraction') of classes of q-vectors at which $\chi_{0}(q,0)$ is maximum. Figure~\ref{fig:009c} displays fluctation maps for La$_{2-x}$Sr$_x$CuO$_4$ (LSCO) (a) and Bi$_2$Sr$_2$CuO$_{6+x}$ (Bi2201) (b), using the full DFT dispersion.  We fix an energy scale by taking $t$ = 0.4195~eV, appropriate for LSCO\cite{RSM_tb}. Most of the colored regions are associated with FS nesting, while the white region is mainly associated with $(\pi,\pi)$ VHS nesting, as will be discussed below.  This figure confirms our earlier finding that LSCO, Fig.~\ref{fig:009c}(a), has a very different phase diagram from most other cuprates, here represented by Bi2201, Fig.~\ref{fig:009c}(b).\cite{Gzm1}  In the phase labeled `ring', the susceptibility is approximately constant along a ring of $q$-vectors surrounding $(\pi,\pi)$.  We note in particular that antinodal nesting (red-shaded region) emerges naturally in most cuprates at higher doping, but is absent in LSCO.  Shown also in Fig.~\ref{fig:009c} are $T_{VHS}$ (dark red-dashed line) and the shift of the DOS-peak (thin green-dashed line), and a series of  mean-field phase boundaries, based on a Stoner criterion, 
\begin{equation}
U\chi_0(q,\omega=0)=1.
\label{eq:0aa}
\end{equation}
There is a critical value of $U=U_0$ below which there is no phase transition at $x=0$.  For $U<U_0$, the phase boundary is dome-shaped, but with a peak shifted away from the $T=0$ value of $x_{VHS}$ to a value close to the line $x_{VHS}(T)$ (red-dashed line).  For $U>U_0$ the phase boundary rapidly evolves to a strong coupling form where the highest $T$ transition occurs near half-filling. Consistent with this, we will show that VHS phenomena are dominated by entropy effects, and thus scale with the $(\pi,\pi)$ susceptibility and not with the peak DOS. In all cases, at sufficiently high $T$, FS nesting is washed out and the dominant instability is at $q=(\pi,\pi)$.  In comparing these phase boundaries to experiments in cuprates, it should be kept in mind that the dotted curves refer to constant $U$, whereas $U$ is likely to be strongly screened near a VHS.\cite{Gonzalez}

\begin{figure}
\leavevmode  
\rotatebox{0}{\scalebox{0.4}{\includegraphics{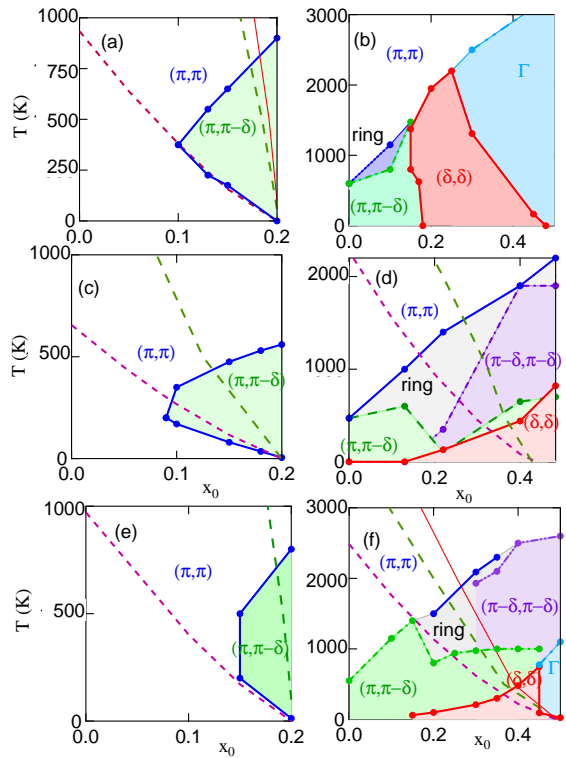}}}
\caption{
{\bf Fluctuation fingerprints of cuprates.}
Maps of $q$-vectors of largest susceptibility for DFT-based models of LSCO (c) and Bi2201 (d) compared to $t-t'-t''$ reference phase a [$t''$ = 0] with $t'$ = -0.23$t$ (a) or -0.43$t$ (b), and reference phase b [$t''=-0.5t'$] with $t'$ = -0.12$t$ (e) or -0.258$t$ (f). Recall, from Fig.~\ref{fig:009c}, that $x_0$ is the doping at $T=0$, but we assume $x\simeq x_0$. }
\label{fig:009} 
\end{figure}

\subsubsection*{Reference families}

The above fluctuation maps allow us to introduce equivalence classes of models for cuprates, and in particular to determine simple reference families that can be used to search parameter space beyond the physical cuprates.  To understand differences between various cuprates, and to gauge their proximity to the pure Hubbard model, it is necessary to add a third axis to the usual $T$- and doping-phase diagrams.  Here we find reference families for which differences between physical cuprates' band structures can be parametrized by a single hopping parameter.

By a Wannier-type downfolding of DFT results, each cuprate can be modelled by a single CuO$_2$ band crossing the Fermi level, so their parameter space consists of hopping parameters, traditionally $t$, $t'$, $t''$, ....  In fact, since $t$ sets an energy scale, the band dispersion is determined by the ratios $t'/t$, $t''/t$, ... To reduce the number of hopping parameters, we introduce a notion of equivalence, wherein a set of materials are equivalent if they have the same fluctuation map. From this equivalence, the DFT parameters can be mapped onto an equivalent {\it reference family} depending only on two parameters, namely $t'/t$ and $t''/t$. We find that when the physical cuprates are approximated by their reference families they lie close to a particular cut in this 2D space. 

Figure~\ref{fig:009} compares the DFT-derived fluctuation maps with similar maps calculated for a variety of simpler dispersions.  The left-hand frames (a), (c), (e), refer to LSCO, the right-hand frames (b), (d), (f), to Bi2201.  The DFT results [middle row, frames (c), (d)] are compared to two reference families, $(i)$ a minimal cut defined by $t''=0$ [top row, (a), (b)] or $(ii)$
 the Pavarini-Andersen cut\cite{PavOK}, $t''=-t'/2$ [bottom row, (e), (f)].  We find that reference family $(ii)$ semi-quantitatively reproduces the DFT derived  fluctuation maps for both cuprates.  On the other hand, setting $t''=0$ works well for LSCO, Fig.~\ref{fig:009}(a), but fails for Bi2201  Fig.~\ref{fig:009}(b). We further find that a reasonable choice of $t'/t$ is the one which matches the Van Hove singularity (VHS) doping of the full DFT dispersion.  We note that both reference families have the additional desirable feature that they evolve from the state with $t'=0$ -- i.e., the original Hubbard model.

\begin{figure}
\leavevmode  
\rotatebox{0}{\scalebox{0.45}{\includegraphics{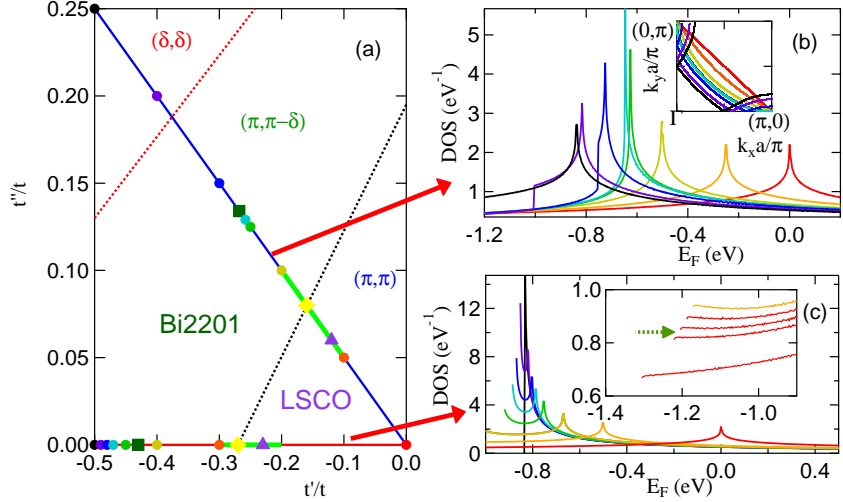}}}
\caption{
{\bf Reference families of the cuprates.} (a) Reference cuts in $t'-t''$-space.  Dotted lines indicate ground state transitions at $x=0$ (these are meant to be  sketches, with only one or two points along the reference lines being determined accurately).  Light green line segments indicate the ranges over which $\xi$ is significantly depressed [Results Section (Beyond RPA)] with maximal depression indicated by yellow diamonds.  Violet triangles [green squares] mark reference states for LSCO [Bi2201]; colored circles indicate $t'$-values for the corresponding DOSs in frame (b) [upper line] or (c) [lower line].  (b)  DOS for several values of $t'$ for reference family $b$ ($t''=-t'/2$).   As VHS moves from right to left, $t'/t$ = 0 (red curve), -0.1 (orange), -0.2 (yellow-green), -0.25 (green), -0.258 (light blue), -0.3 (blue), -0.4 (violet),  and -0.5 (black).   Inset: corresponding Fermi surfaces at VHS.  (c)  DOS for several values of $t'$ for reference family $a$ ($t''=0$).  As VHS moves from right to left, $t'/t$ = 0 (red curve), -0.3 (orange), -0.4 (yellow-green), -0.45 (green), -0.47 (light blue), -0.48 (blue), -0.49 (violet), and -0.5 (black). Inset: leading edge of DOS, near band bottom for (from bottom to top) $t'/t$ = -0.22, -0.27, -0.28, -0.29, -0.30, in regime where it crosses over from a local minimum to a local maximum.
} 
\label{fig:009d} 
\end{figure}
An advantage of using the reference families is that this lets us systematically vary parameter space to study the evolution from one cuprate to another, or to extrapolate the parameter space to better understand the origin of various phases.  Figure~\ref{fig:009d}(a) shows the cuts in $t'-t''$-space associated with reference families $(i)$ ($t''=0$, red line) and $(ii)$ ($t''=-t'/2$, blue line).  The dotted lines, Figure~\ref{fig:009d}(a), indicate ground-state transitions at $x=0$, from $(\pi,\pi)$ to $(\pi,\pi-\delta)$ (black dots) and from $(\pi,\pi-\delta)$ to $(\delta,\delta)$ order (red dots).  The black dotted line is the Mott-Slater line discussed in the Results Section (Beyond RPA).  The densities-of-states (DOS) shown for the two families in Figs.~\ref{fig:009d}(b) and~\ref{fig:009d}(c) provide insight into the origin of the crossovers.  Note that the use of these reference cuts demonstrates the clear evolution of the VHS between 2D and 1D behavior.

For reference family $(i)$, Fig.~\ref{fig:009d}(c), the DOS reveals a crossover associated with a change from 2D to 1D nesting. This is seen by the DOS at $t'=-t/2$ which shows the characteristic $1/\sqrt{E}$ divergence expected for a 1D system.  To explore the origin of the 1D-behavior we consider the electronic dispersion for reference family $(i)$ of the form $E=-2t(\cos(k_x a)+\cos(k_y a))-4t'\cos(k_x a)\cos(k_y a)$. If either $k_x$ or $k_y$ = 0 and $t'=-t/2$ then $E= -2t$. Therefore the FS at the VHS energy reduces to just the $x$ and $y$ axes. This shows that the strong 1D nature of the corresponding FS leads to a strong divergence in the DOS at the VHS. 

Near $t'=-t/2$, the DOS has two peaks, one at the saddle-point VHS found in 2D materials, the other at the leading edge, as found in 1D systems, Fig.~\ref{fig:009d}(c) (violet line).  For $t'<-0.45t$ (green-black), the leading edge is larger, suggestive of weakly-coupled chains, while for $t'>-0.28t$ (green dotted arrow) the leading edge peak turns into a local minimum, Fig.~\ref{fig:009d}(c) inset, at the green dotted arrow.  This quasi-1D behavior is suggestive of the nematic phase found in cuprates.\cite{nema2}  Indeed, an earlier VHS renormalization group calculation found that charge order transitions splitting the degeneracy of the VHSs at $(0,\pi)$ and $(\pi,0)$ [which would now be called a nematic transition] are only possible if $t'/t<-0.276$.\cite{Gonzalez}  
Remarkably, $t'=-0.28t$ approximately coincides with the commensurate-incommensurate transition found in Figs.~\ref{fig:3g2N}(b) 
to lead to a severe $\xi$ depression.

For family $(ii)$ a non-zero $t''$ modifies the dispersion such that it breaks up the 1 dimensionality of the FS and weakens the DOS divergence, as seen in Fig.~\ref{fig:009d}(b). However, a residual 1 dimensionality remains in that there is conventional FS nesting in the ANN region, which is responsible for the CDW. For the red, orange, yellow, and green curves we find a resemblance to that of the DOS for $t''=0$. At the light blue curve there is a topological transition where a pocket centered at $(\pi,0)$ splits off from the rest of the Fermi surface. For larger $t'$, one now sees two features in the DOS, a peak followed by a step at lower energies. The peak is the topological transition where the pockets split off, and the step is where the $(\pi,0)$ pocket disappears. The inset illustrates this progression showing the FS at the DOS peak energy for various $t'$.

\subsection*{Origins of structure in the susceptibility: competition between FS and VHS nesting}\label{sec:3b}
\subsubsection*{Separating FS maps and broad background}

In order to understand the structure in the fluctuation maps we need to analyze the susceptibilities more deeply. The susceptibility contains sharp features which are a map of the Fermi surface and which quantify the strength of the Fermi surface nesting. In addition, the susceptibility contains a smooth, intense background feature which peaks at $(\pi,\pi)$.  This peak plays an important role in the cuprates, producing large entropic effects.  This peak can shift the balance of the FS nesting to $q$-vectors closer to the peaks, and in special cases, it can lead to commensurate nesting away from the FS nesting vector at exactly $(\pi,\pi)$.  Moreover, as $T$ increases, coherent FS features are washed out, leaving behind only the commensurate bulk contribution.  
Here we determine that this background can be considered to be a bosonic VHS (b-VHS), in that the electronic excitation comes predominantly from the region of the VHS.  It is analogous to the Van Hove excitons found in optical spectra\cite{33Phil}, but present in the intraband susceptibility. The evolution of this feature contains hints as to the origin of the pseudogap. 

Consider the expression for the $\omega=0$ bare susceptibility that enters into the Stoner criterion,
\begin{equation}
\chi_0(q)=\sum_k\frac{\Delta f_{k,q}}{\epsilon_k-\epsilon_{k+q}},
\label{eq:0a4}
\end{equation}
where $\Delta f_{k,q}$ is the Pauli blocking factor
\begin{equation}
\Delta f_{k,q}=f(\epsilon_{k+q})-f(\epsilon_k).
\label{eq:0a5}
\end{equation}
The denominator is smallest if $\epsilon_k=\epsilon_{k+q}$, but in this case the numerator vanishes at $T=0$ unless $\epsilon_k$ is at the Fermi level.  Even in this case, the term contributes little weight to the susceptibility integral unless the $q$-shifted FSs are tangential, which is generally satisfied if $q=2k_F$, giving rise to a folded map of the FS -- the ridge of FS nesting\cite{Gzm1}.

\begin{figure}
\leavevmode  
\epsfig{file=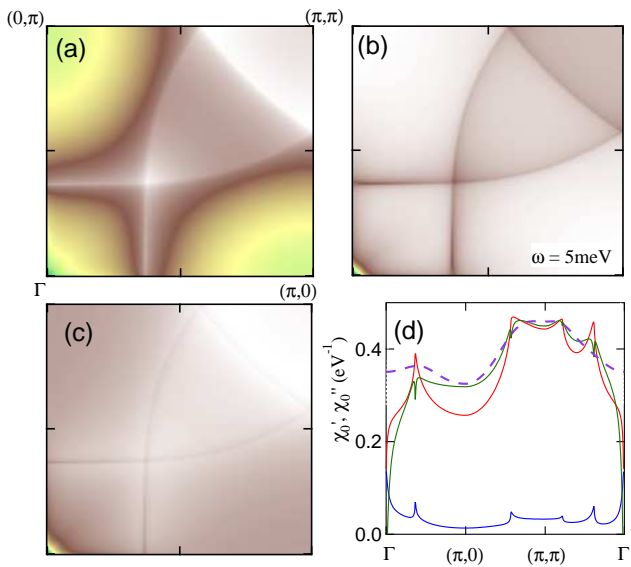,width=9.5cm, angle=0}
\vskip0.5cm
\caption{{\bf Extracting the bulk $\chi_0$ contribution.}
(a) Susceptibility $\chi'_0$ of Bi2201 for full DFT dispersion and $x=0.13$, plotted in first Brillouin zone at $T$ = 0, $\omega=$~5~meV.  Color scale from minimum (blue) to maximum (white). 
(b)  As in (a), but for $\chi''_0$.  (c) Difference plot of (a) and the scaled version of (b).
(d) Cuts of susceptibility from earlier frames plotted along high symmetry lines, with red curve representing data in frame (a), blue curve for frame (b), and green curve for frame (c).  Shown also is the curve corresponding to frame (a), but at $T=1000$K (violet long-dashed line) scaled by 1.4 and shifted by -0.045~eV$^{-1}$ to match other curves near $(\pi,\pi)$.
} 
\label{fig:01} 
\end{figure}

The Fermi surface nesting features in the susceptibility map for Bi$_2$Sr$_2$CuO$_6$ (Bi2201), at $x=0.13$ and $\omega = 5$~meV, are illustrated in Fig.~\ref{fig:01}, with $\chi_0'$ given in frame (a), and the corresponding map of $\chi_0''$ in frame (b). Due to the energy $\delta$-function, $\chi_0''$ contains only the FS contribution\cite{Mazin}, except an extra feature at $\Gamma$ associated with the DOS.  In Fig.~\ref{fig:01}(d), we plot $\chi_0'$ (red line) and $\chi_0''$ (blue line) along a cut $\Gamma\rightarrow (\pi,0)\rightarrow (\pi,\pi)\rightarrow\Gamma$ in momentum space.  It can be seen that, away from $\Gamma$, the peaks in both components fall at the same $q$-values, and the relative weights of the peaks are comparable.  Moreover, the anti-nodal nesting peak near $\Gamma$ in $\chi_0''$ is stronger than the near-$(\pi,\pi)$ peak.  However, $\chi_0'$ is largest on the near-$(\pi,\pi)$ peaks, since the FS contribution is riding on the large background contribution to the susceptibility.  Here we demonstrate that VHS nesting plays a dominant role in creating this background, thereby providing a deeper understanding of the VHS vs FS nesting competition.

Two approaches are taken to eliminate the FS nesting features to see the background susceptibility by itself.  First, raising $T$ (violet long-dashed line in Fig.~\ref{fig:01}(d)) washes out the sharp FS features first, while leaving the background features behind.  
Secondly, in frame (c), we plot the difference between $\chi_0'$ and a suitably scaled quantity proportional to $\chi_0''$; this is also plotted as the green curve in frame (d).  Since $\chi_0'$ and $\chi_0''$ are Kramers-Kr\"onig transforms, they have different lineshapes, but since the FS map is nearly one-dimensional, a large part of it is still cancelled in taking the difference.  From frame (d) it can be seen that both approaches lead to similar results for the bulk susceptibility, and that this contribution provides the dominant contribution to virtually the whole $(\pi,\pi)$ plateau structure. Here the background contribution arises as follows. Even when $\epsilon_k\ne\epsilon_{k+q}$, there will be a finite contribution to the susceptibility, Eq.~\ref{eq:0a4}, reduced by the factor $1/(\epsilon_k-\epsilon_{k+q})$, as long as the two energies lie on opposite sides of the Fermi level (for $T=0$).  This contribution will be especially significant near a VHS, which has a large spectral weight.

\subsubsection*{Background peak as VHS nesting}

\begin{figure}
\leavevmode
\rotatebox{0}{\scalebox{1.2}{\includegraphics{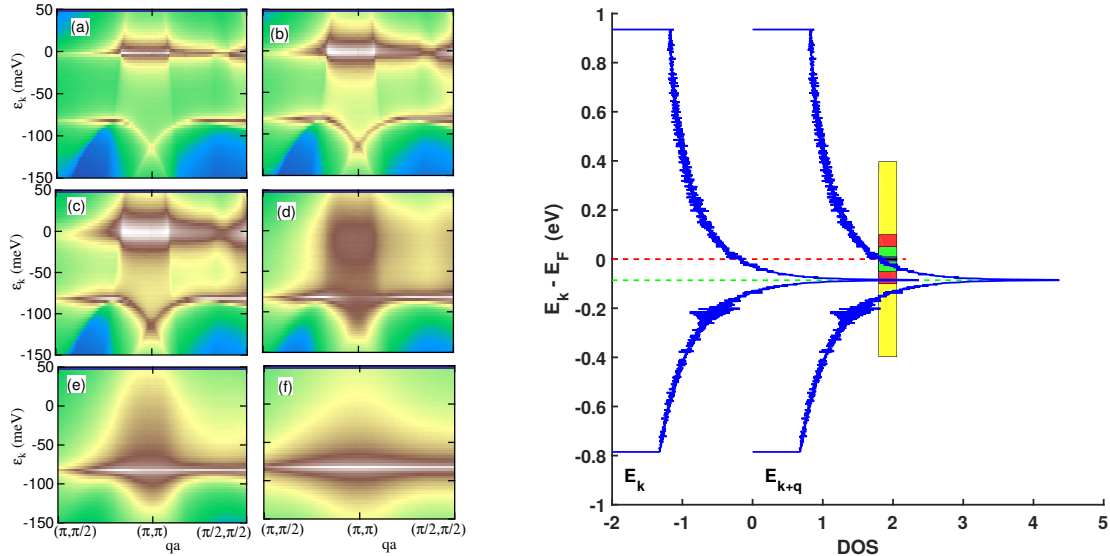}}}
\vskip0.5cm  
\caption{
{\bf Coherent-incoherent crossover at x=0: role of VHS nesting.}
Deconvolution of the susceptibility, $\chi^{\prime}_{0}$, for $k$-states in the $(\pi,0)$-patch at several temperatures $T$.  Contribution to the susceptibility of $\chi_{0,2}(q,\epsilon_k)$, Eq.~\ref{eq:0a4b}, is plotted along $q$-cuts near $(\pi,\pi)$ with the color scale ranging from white for a large contribution to blue for negligible contribution for: $T$= 0 (a), 25K (b), 50K (c), 250K (d), 500K (e), and 2000K (f).  Frame (g) illustrates scattering from the VHS at $E_k$ (green dotted line) to states centered at the Fermi level at $E_{k+q}$. The colored region shows where Pauli blocking is ineffective at $T$ = 25 (blue bar), 50 (black bar), 250 (green bar), 500 (red bar), and 2000K (yellow bar).
}
\label{fig:6}
\end{figure}

Here we demonstrate the connection between the bulk contribution to $\chi_0$ and VHS nesting by deconvolving the bare susceptibility, Eq.~\ref{eq:0a4}, into its various $\epsilon_k$ components.  To understand this procedure, recall that the VHS  gives rise to susceptibility features near $\Gamma$ (DOS) and $(\pi,\pi)$. The latter are typically more intense. Thus, to confirm that the background susceptibility is associated with the VHS, we analyze Eq.~\ref{eq:0a4} for $\chi_0$, assuming that the electron is associated with $\epsilon_k$ and the hole with $\epsilon_{k+q}$.  We must demonstrate that, when $q\sim (\pi,\pi)$, the dominant contribution to $\chi_0$ has $k$ near the VHS at $(\pi,0)$ [or $(0,\pi)$] and $\epsilon_k$ near $E_{VHS}$. 

\begin{figure}
\leavevmode
\rotatebox{0}{\scalebox{0.84}{\includegraphics{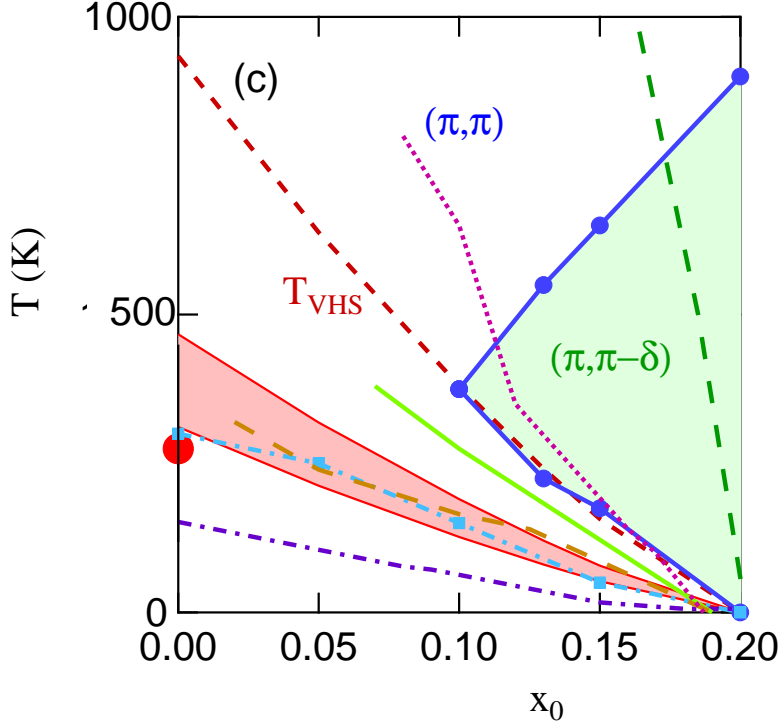}}}
\vskip0.5cm  
\caption{
{\bf LSCO phase diagram, showing several VHS-related crossovers
and regions where the susceptibility peaks at different $q$-vectors}. Dominant fluctuations are at commensurate $(\pi,\pi)$ (white shaded region) or incommensurate $(\pi,\pi-\delta)$ (green shaded region); crossovers are $T_{VHS}$ (red short-dashed line), DOS peak (green long-dashed line), coherent-incoherent crossover (pink shaded region), $T_{\gamma}$ (violet dot-dashed curve), and the position of the $(\pi,\pi)$ peak vs doping (light blue dot-dashed line).
Experimental pseudogaps from Refs.~\onlinecite{9Doir} (yellow-green solid line and brown dashed line) and~\onlinecite{Tall} (magenta dotted line).  Recall, from Fig.~\ref{fig:009c}, that $x_0$ is the doping at $T=0$, but we assume $x\simeq x_0$. 
}
\label{fig:2c}
\end{figure}

To demonstrate this we divide the Brillouin zone into four patches, around the $\Gamma$, $(\pi,0)$, $(0,\pi)$, and $(\pi,\pi)$ points, and calculate separately the susceptibility due to each patch, rewriting Eq.~\ref{eq:0a4} as
\begin{equation}
\chi_0(q)=\sum_{i=1,4}\sum_{\epsilon_k}\chi_{0,i}(q,\epsilon_k),
\label{eq:0a4b}
\end{equation}
where $i$ runs over the four quadrants $\Gamma$, $(\pi,0)$, $(0,\pi)$, and $(\pi,\pi)$.  In LSCO, the plateau is predominantly composed of equal contributions from the $(0,\pi)$ and $(\pi,0)$ patches, as expected for a VHS.  We then explore the distribution of values of $\epsilon_k$ which contribute to the susceptibility in the $(\pi,0)$ patch at $x=0$, Fig.~\ref{fig:6}.  
There is a clear crossover from a dominant near-FS contribution ($\epsilon_k=0$) at low $T$ to a dominant contribution at high-$T$ which lies along a flat (q-independent) line centered at energy $k_BT_{VHS}=E_F-E_{VHS}$. As we vary doping, this flat line shifts towards the Fermi level following $T_{VHS}$, red-dashed line in Figs.~\ref{fig:009}(a),~\ref{fig:2c}.

This crossover is associated with Pauli blocking, i.e. the factor $\Delta f$ in Eq.~\ref{eq:0a5}. Since at low $T$ the Fermi function is a step, $\Delta f$ blocks the scattering between states far below $E_F$.  As $T$ increases $f$ gets smeared out, so that features away from the Fermi surface can contribute to the susceptibility. Since the VHS has a substantial peak, the high-$T$ changes are due to the unveiling of this feature.  Note that at low $T$, Figs.~\ref{fig:6}(a-c), there is still scattering from one VHS, but to states near $E_F$, and hence contributing intensity at $q\ne (\pi,\pi)$.  Correspondingly, the weight near $(\pi,\pi)$ comes from states in the tail of the VHS extending to $E_F$.

While $\Delta f$ enhances features associated with the VHS as $T$ increases, it has the opposite effect on features near the Fermi level.  For energy  $\epsilon$ exactly at $E_F$, $f(\epsilon)$ is exactly 1/2 at any $T$, so $\Delta f$ =0 if both states are at $E_F$.  As $T$ increases, $f$ smooths out, so that if $|\epsilon_{\bf k}|$, $|\epsilon_{\bf k+q}|$ are both $<k_{B}T$, then $f_{\bf k}\approx f_{\bf k+q}\approx 1/2$, and therefore, $\Delta f \approx 0$.  This explains why the crossover from FS nesting to VHS nesting in Fig.~\ref{fig:6} occurs in such a narrow $T$ range (pink shaded region in Fig.~\ref{fig:2c}). Passing between Figs.~\ref{fig:6}(c),~(d) the Fermi surface contribution goes through a coherent - incoherent crossover -- note that the FS nesting contribution is fully incoherent by 500K.

\begin{figure}
\leavevmode
\rotatebox{0}{\scalebox{0.38}{\includegraphics{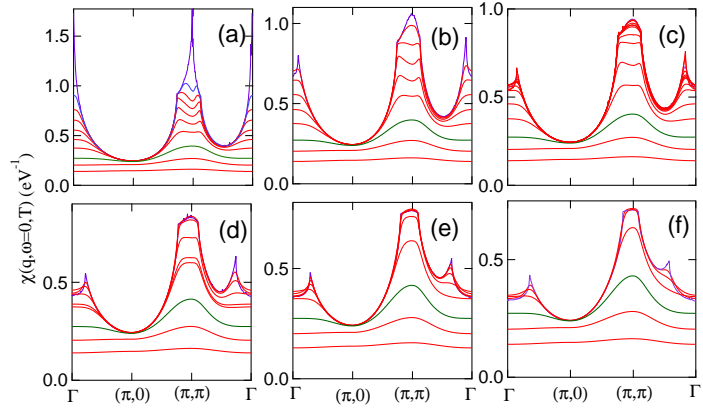}}}
\vskip0.5cm  
\caption{
{\bf Coherent-incoherent crossover at x=0 in $\chi_0(T)$.}
Lindhart susceptibility $\chi'(q,\omega=0)$ of LSCO, plotted along high-symmetry lines at 6 different dopings:
(a) $x=0.20$ ($\sim$VHS), (b) $x$ = 0.18, (c) $x$ = 0.15, (d) $x$ = 0.10, (e) $x$ = 0.05, and  (f) $x$ = 0, 
at (from top to bottom) $T$ = 0, 10, 25, 50, 100, 200, 400, 1000, 2000, and 4000K, except that in frames (d) and (f) the traces at $T$ = 10 and 25K are omitted.  The $T=1000K$ susceptibility is plotted as a green line. 
}
\label{fig:6z}
\end{figure}

  In  Fig.~\ref{fig:2c} we superpose this crossover, plotted as a pink shaded region, on the fluctuation map for LSCO, with all calculations for the minimal reference family ($t''=0$). This incoherent-coherent crossover scales with $T_{VHS}$, with boundary lines given by $T_{VHS}/2$ and $T_{VHS}/3$.  The same crossover can be seen directly in the susceptibility, Fig.~\ref{fig:6z}, as the coherent, FS-related features present at low $T$ are washed out, leaving only a featureless (incoherent) $(\pi,\pi)$ plateau at high $T$.

These effects of Pauli blocking can explain several additional features of  Fig.~\ref{fig:2c}.  For instance, hole doping causes the $(\pi,\pi)$-plateau to expand in momentum space, leading to a weakening of VHS nesting.  Thus, in Bi2201, the transition between $(\pi,\pi)$-nesting and some form of FS nesting shifts to higher $T$ with increasing $x$, as seen in Figs.~\ref{fig:009}(b,d,f).  A similar effect is found in LSCO, Figs.~\ref{fig:009}(a,c,e), for the high-$T$ termination of the incommensurate $(\pi,\pi-\delta)$ order (green shaded region). However, in LSCO there is also an anomalous reentrant transition, where $(\pi,\pi)$ order reappears below a lower domain boundary (blue solid line).  
 To understand this unusual reentrance, consider first what happens to $\chi_0$ when $E_F=E_{VHS}$, Fig.~\ref{fig:6z}(a). While the VHS peak is clearly present at $T=0$ (blue line), finite temperature does not lead to simple thermal broadening but actually splits the peak. For small finite $T$ the very sharp VHS at $E_F$ is blocked. As T increases, the blocking spreads over a wider range of energy, leading to the splitting of the VHS feature. Thus, raising $T$ causes a blocking first of the sharp VHS peak to produce the lower edge of the domain boundary.  Shifting to $x<x_{VHS}$ means that a higher $T$ is necessary to block the sharp VHS peak.  Note that the lower branch of the boundary closely follows $T_{VHS}$, especially for the $t''=0$ reference family.  This is also why the conventional VHS theory that relies on the sharp VHS peak only holds at relatively low $T$ as discussed in conjunction with  Fig.~\ref{fig:009c}.


\subsubsection*{Pauli-blocking and the bosonic-VHS}

\begin{figure}
\leavevmode
\rotatebox{0}{\scalebox{0.34}{\includegraphics{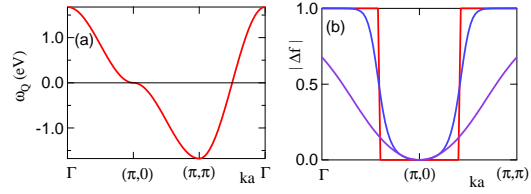}}}
\vskip0.5cm  
\caption{
{\bf Coherent-incoherent crossover at the VHS.}
(a) Pair dispersion, $\omega_Q$, as a function of $k$, and (b) the corresponding weight $\Delta f$, at $T$ = 0K (red), 500K (blue), and 2000K (violet) for LSCO ( $x=0$).
}
\label{fig:2cc}
\end{figure}

If we think of $\chi_0$ as the propagator of different electron-hole pairs, then the pair associated with $q=Q\equiv (\pi,\pi)$ has many similarities with the excitonic-VHS seen in optical studies\cite{33Phil}, and we will refer to it as the bosonic-VHS, or b-VHS for short.  We clarify  the role of Pauli blocking by describing its effect on the b-VHS in detail, using the $t''=0$ reference family for simplicity.  To understand the relationship between the b-VHS and entropy effects, we need to know how Pauli blocking ``hides'' the states associated with the VHS at low $T$, and how they recover as $T$ increases.  If the renormalized dispersion of a single electron is $\epsilon_k$ with wave vector $k$, then an e-h pair at wave vector $q$ has a dispersion $\omega_q(k)=\epsilon_{k+q}-\epsilon_k=-2\epsilon_{q-}(k)$, where $\epsilon_{q\pm}(k)=(\epsilon_k\pm\epsilon_{k+q})/2$, and a Pauli blocking factor $\Delta f_{k,q}=f(\epsilon_{k+q})-f(\epsilon_k)$.  Then the corresponding pair DOS is $D_q(\omega)=\sum_k\Delta f_{k,q}\delta(\omega- \omega_q(k))=\chi''_0(q,\omega)$.  For LSCO, the dominant pairs are those at $q=Q$, the AF nesting vector.   The associated dispersion $\omega_Q(\pi,\pi)$, plotted in Fig.~\ref{fig:2cc}(a), resembles the electronic dispersion $\epsilon_k$, but with an important distinction: it depends only on $\epsilon_{Q-}(k)$, whereas all of the hopping terms that shift the electronic VHS away from half-filling ($t'$, $t''$) are contained in $\epsilon_{Q+}(k)$, i.e. the b-VHS is pinned at $\omega=0$.  Since $\chi_0''$ is an odd function of $\omega$, $\chi''_0(Q,\omega =0)=0$.   However, while the b-VHS is pinned at $\omega=0$, Fig.~\ref{fig:2cc}(a), its weight vanishes at $T=0$, due to the Pauli-blocking factor, $\Delta f _{k,Q}=0$ near $k=(\pi,0)$ at $T=0$, Fig.~\ref{fig:2cc}(b). Finite $T$ restores weight, optimally near 1000K, although $\Delta f$ always vanishes exactly at $(\pi,0)$.  Hence the bosonic entropy only turns on at finite $T$.  Figure~\ref{fig:6}(g) illustrates how the Pauli window opens with increasing $T$.

 As $T$ increases, the Fermi functions spread out, so the FS contribution decreases whereas the bulk contribution grows as $\Delta f$ becomes unblocked.  For the VHS, we expect full unblocking at a temperature near $T\sim T_{VHS}$ (red short-dashed line in Fig.~\ref{fig:2c}).  Since $f$ varies exponentially with $T$, the unblocking will appear at a temperature $T<T_{VHS}$, which may vary according to the property studied.   This is similar to a Schottky anomaly, where a feature at energy $\Delta$ above the ground state produces a peak in the heat capacity at a temperature $T_m\sim\Delta /2.4$.
The low-$T$ fadeout of the b-VHS is controlled by the Pauli blocking factor, 
\begin{equation}
\Delta f_{k,Q}=\frac{\sinh{x_-}}{\cosh{x_-}+\cosh{x_+}},
\label{eq:0a5}
\end{equation}
where $x_-=\epsilon_{Q,-}(k)/k_BT$, $x_+=(\epsilon_{Q,+}(k)-E_F)/k_BT$. When the electronic VHS is at the Fermi level, $x_+=0$ and $\Delta f_{k,Q}=\tanh{x_-/2}$, while at lower doping the b-VHS is exponentially suppressed, $\cosh{x_+}\sim exp(T_{VHS}/T)/2$, which equals 1 when $T\sim T_{VHS}/ln(2)$.

\subsubsection*{Entropy and a revised VHS scenario}

\begin{figure}
\leavevmode
\rotatebox{0}{\scalebox{0.38}{\includegraphics{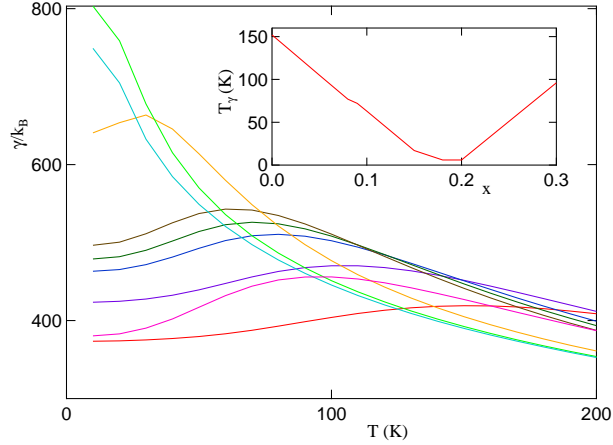}}}
\vskip0.5cm  
\caption{
{\bf Sommerfeld constant calculated at several dopings.} $x$ = 0 (red curve), 0.05 (violet), 0.08 (blue), 0.09 (green), 0.10 (brown), 0.15 (orange), 0.18 (light green), 0.20 (light blue), and 0.30 (magenta).  Inset: $T_{\gamma}$ vs doping.
}
\label{fig:2d}
\end{figure}

 Here we will demonstrate that this b-VHS peak produces large entropic effects. 
To show this, we calculate several characteristic VHS-dominated properties -- the Sommerfeld coefficient of the heat capacity and the susceptibilities at $\Gamma$ and $(\pi,\pi)$. The entropy is calculated from the standard expression
\begin{equation}
S=-k_B\sum_k\bigl[f_kln(f_k)+(1-f_k)ln(1-f_k)\bigr], 
\label{eq:0a6}
\end{equation}
with $\gamma=dS/dT$, Fig.~\ref{fig:2d}.  The sum is over the electronic dispersion of LSCO, assuming a paramagnetic phase.  At $T=0$, $\gamma$ is proportional to the DOS, and hence diverges at $x_{VHS}$, Fig.~\ref{fig:2d}.   However, as $T$ increases, the peak does not remain near $x_{VHS}$, but rapidly shifts toward $x=0$.    We plot the peak position $T_{\gamma}$ both in the inset of Fig.~\ref{fig:2d} and as a violet dot-dashed line in Fig.~\ref{fig:2c}.  Clearly, since $\gamma$ is a measure of entropy, it is sensitive to sources of entropy away from the Fermi level, when $T$ is high enough to reduce the Pauli blocking.  It has recently been noted that there should be an excess of low-$T$ entropy in the vicinity of a quantum phase transition.\cite{34Canf}

Similar behavior is found for $\chi_0$.  The VHS has logarithmically diverging features in the susceptibility at two different $q$-vectors, an inter-VHS contribution at $(\pi,\pi)$, and an intra-VHS contribution at $\Gamma$ \cite{VHS1,VHS2}. While both peaks cross the Fermi level at $x_{VHS}$ at $T=0$, they evolve differently with temperature.  This is because the feature at $\Gamma$ represents the peak in the  DOS, and hence it is dominated by near-FS physics.  As a result it shifts only weakly with $T$, green long-dashed line in Fig.~\ref{fig:2c}.  In contrast, the inter-VHS $(\pi,\pi)$ peak is a bulk contribution and hence is sensitive to Pauli blocking, as discussed above. This leads to the rapid thermal evolution of the $(\pi,\pi)$-susceptibility peak all the way to $x=0$ (light blue dot-dashed line), scaling with both $T_{\gamma}$ and $T_{VHS}$, Fig.~\ref{fig:2c}.  However, this behavior seems to be dominated by a separate aspect of the VHS physics.  As discussed in Ref.~\onlinecite{Gzm1}, the susceptibility has an approximate electron-hole symmetry, with similar spectra for a nearly empty ($x\sim 1$) or a nearly full ($x\sim -1$) band.  The $t'$ and $t''$ hopping parameters break this symmetry, and as a result, the low-$T$ susceptibility is roughly symmetrical about $x_{VHS}$, where it has its largest peak.  Here it seems that as $T$ increases, the electron-hole symmetry is restored when $T>t'$.  Since $T_{VHS}\sim t'$, one expects a similar scaling of $T$ vs $x$ for $x\ge 0$.  But when $x<0$ $T_{VHS}$ continues to increase, whereas the peak in $\chi_0(q=(\pi,\pi))$ stays at $x=0$, indicative of a restored electron-hole symmetry (see Supplementary Material Fig.~2).

These entropic effects have profound consequences for VHS physics that were overlooked in earlier mean-field studies.\cite{VHS1,VHS2}
Most earlier studies of cuprate antiferromagnetism used a model valid only near the FS, or at low-$T$, overlooking Pauli unblocking of the VHS feature at large $T$.  As a consequence, they predicted that $\chi_0$ would show a maximum for $x$ near $x_{VHS}$, inconsistent with experiment. Figure~\ref{fig:009c}(a) shows revised RPA phase diagrams for several values of $U$, based on the DFT-Lindhard susceptibility. For small enough $U$ ($U<\sim$~1.3~eV), the earlier results are recovered, but when $U$ is larger a new behavior is found, with a maximum transition temperature at half-filling, decreasing with doping as if approaching a quantum critical point.  This now starts to resemble the pseudogap phase diagram of the cuprates. The  comparison of experimental pseudogap lines\cite{9Doir,Tall} in Fig.~\ref{fig:2c} will be discussed in the Discussion Section.

Finally, when McMillan\cite{25McMi} introduced the idea of bosonic entropy, he meant that competition between different phonon instabilities (phonons with different $q$-vectors) slowed down the tendency to CDW ordering, leading to broad $T$-ranges with only short-range order.  In the following section we will show that a similar competition exists for electronic bosons, and that mode coupling allows us to quantify these effects.

\section*{Results: Beyond RPA susceptibility}

\subsection*{Mode coupling formalism}
Above, we have established the outlines of a classification scheme for phase transitions, analogous to the spectrum-generating algebras (SGAs) of nuclear physics.  We summarize this development in Supplementary Material Section I.A, and compare it to SGAs in Supplementary Material Section I.C. In the present section we show how competition between condensing phases affects the transition temperature, and how to quantify the number of competing modes. A few striking examples of this competition will be discussed.

From $\chi_0$ we calculate an interacting $\chi$ of modified random-phase approximation (RPA) form,
\begin{equation}
\chi ({\bf q},i\omega_n)={\chi_0({\bf q},i\omega_n)\over 1+\lambda 
-U\chi_0({\bf q},i\omega_n)},
\label{eq:1b} 
\end{equation}
where $\lambda$ is a mode-coupling parameter which must be calculated self-consistently.  The imaginary part of $\chi$ can be thought of as the density-of-states (DOS) of electronic bosons, electron-hole (e-h) pairs, which may become excitons or excitonic resonances when a long-range Coulomb interaction is turned on.  However, static ($\omega =0)$  instabilities depend on a Stoner criterion, and hence on
\begin{equation}
\chi'_0(q,\omega=0)=\sum_k\frac{f(\epsilon_{k+q})-f(\epsilon_k)}{\epsilon_k-\epsilon_{k+q}}
=2\int_0^{\infty} \frac{d\omega'}{\pi} \frac{\chi''_0(q,\omega')}{\omega'}.
\label{eq:E2}
\end{equation}


The self-consistent parameter $\lambda$ is found from a Matsubara sum of the susceptibility
\begin{equation}
\lambda ={A_0T\over N}\sum_{{\bf q},i\omega_n}\chi ({\bf q},i\omega_n),
\label{eq:1}
\end{equation}
where $N$ is the number of $q$-points, and the summation in Eq.~\ref{eq:1} can be transformed:
\begin{eqnarray}
{T\over N}\sum_{{\bf q},i\omega_n}\chi ({\bf q},i\omega_n)&=&\int 
{d^2qa^2\over 4\pi^2} \int_{0}^{\infty}{d\omega\over\pi}coth({\omega\over 
2T})\chi''({\bf q},\omega +i\delta )
\nonumber \\
&\simeq&\hat\lambda_1+\hat\lambda_2T,
\label{eq:2}
\end{eqnarray}
with $a$ the in-plane lattice constant,
\begin{equation}
\hat\lambda_1=\int {d^2qa^2\over 4\pi^2}
\int_{0}^{\infty}{d\omega\over\pi}\chi''({\bf q},\omega),
\label{eq:2b}
\end{equation}
\begin{equation}
\hat\lambda_2=\int {d^2qa^2\over 4\pi^2}\chi' ({\bf q},0 ).
\label{eq:2c}
\end{equation}
The term 
$\hat\lambda_1$ introduces a small, nonsingular correction\cite{23RSMMC} to 
Eq.~\ref{eq:1}, which we neglect.  

\begin{figure}
\leavevmode
\rotatebox{0}{\scalebox{0.38}{\includegraphics{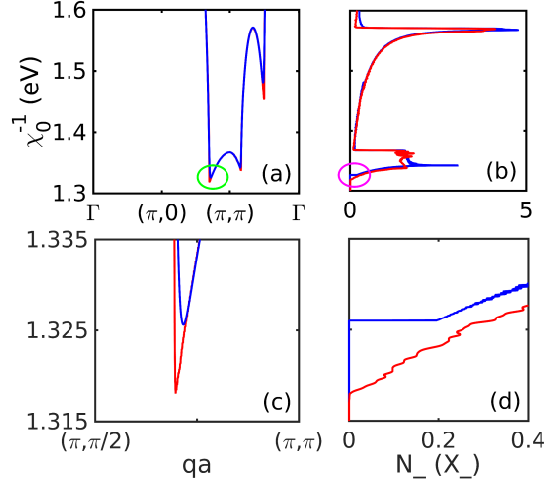}}}
\vskip0.5cm  
\caption{
{\bf Origin of SDOS features.}
(a) Inverse susceptibility $\chi_0^{-1}({\bf q})$ for Bi2201  at doping
$x=0$.  Note that $U_c=min(\chi_0 '^{-1})$ (circled region). Curves are at $T$ = 10 (red) and 100K (blue).  (b) Corresponding susceptibility density of states (SDOS) $N_-(X_-)$, plotted horizontally, corresponding to the susceptibility of (a).  
(c,d) Blowups of circled regions in (a,b).
}
\label{fig:2b}
\end{figure}

In order to explore the role of suseptibility plateaus for realistic band dispersions, the self-consistency equation is evaluated numerically. 
Note that since $\chi_0^{-1}$ has dimensions of energy, a plot of $\chi_0^{-1}(q,0)$ resembles a dispersion map.  Hence we can define a SDOS: 
\begin{eqnarray}
\int {d^2qa^2\over 4\pi^2}=\int N_-(X_-) dX_-,
\label{eq:6}
\end{eqnarray}
where $X_-=\chi_0^{-1}$, is the variable of integration, and we introduce a corresponding susceptibility density of states (SDOS) $N_-$. Eq.~\ref{eq:1} then becomes 
\begin{eqnarray}
\lambda =\Gamma TA_0\int_{U_c}^{\infty} dX_- { N_-(X_-)\over X_--U_c+\delta}, 
\label{eq:5}
\end{eqnarray}
with $\delta = U_c-U_{sp}$, $U_c= 1/\chi_0({\bf Q_0},0)$.
In Section II of Supplementary Materials, we show that Eq.~\ref{eq:5} is closely related to excitonic Bose condensation, with $\lambda$ proportional to the effective number of bosons.
Once $N_-$ has been calculated, Eq.~\ref{eq:5} can be evaluated numerically.  The singular part of the integral is treated analytically, and the remainder numerically, with  $u$ and $U$ approximated as constants, $u=0.8eV^{-1}$, $U$ = 2~eV. It is convenient to fix the SDOS at some temperature $T'$, and then solve Eq.~\ref{eq:5} for $T(\xi_{th} ,T')$, with self-consistency requiring $T(\xi_{th} ,T')=T'$.

Figure~\ref{fig:2b} illustrates how integrating over the inverse susceptibility, Fig.~\ref{fig:2b}(a), leads to the SDOS, Fig.~\ref{fig:2b}(b), at $x=0$ where the near-$(\pi,\pi)$ plateau is dominant.  
Here we use hopping parameters appropriate to the DFT dispersion of Bi2201, $t=419.5$, $t'=-108.2$ and $t''=54.1$~meV (Subsection~Mode coupling in Bi2201).
This yields a doping phase diagram  that is qualitatively similar to that of most cuprates, except LSCO.  By comparing Figs.~\ref{fig:2b}(a) and~\ref{fig:2b}(b), one can see how features in $\chi_0^{-1}$ translate into features in $N_-$.  Thus, the intense, flat-topped peak in $N_-$ at small values of $\chi_0^{-1}$ represents the near-$(\pi,\pi)$ plateau.  Its broad leading edge (smaller $\chi_0^{-1}$) is controlled by anisotropy of the plateau edge between $(\pi,\pi-\delta)$ and $(\pi-\delta,\pi-\delta)$, while the sharp trailing edge corresponds to the local maximum of $\chi_0^{-1}$ at $(\pi,\pi)$.  For the ANN peak, its leading edge scarcely leaves any feature in $N_-$, but a local maximum translates into a large peak in $N_-$.  Finite temperature, $T=100K$ (blue line), rounds off the cusp in $\chi_0^{-1}$, Fig.~\ref{fig:2b}(c), leading to a step in $N_-$, Fig.~\ref{fig:2b}(d), but otherwise has little effect.

Instead of such a self-consistent calculation, many treatments of phase transitions, including path-integral approaches, spin-fermion models, and Hertz-Millis\cite{35Hert,36Mill} quantum critical theories, approximate the susceptibility denominator by a $T$-independent Ornstein-Zernicke(OZ)\cite{37Orns} form, 
\begin{eqnarray}
\chi ({\bf q},\omega )\sim{1\over \bar q^2+\xi^{-2}+i(\omega /\omega_c)^z},
\label{eq:3}
\end{eqnarray}
in terms of various deviations from the critical point (in $\bar q$, $\omega$, and a `tuning parameter' which is proportional to $\xi^{-2}$, where $\xi$ is the correlation length).  Here $z$ is a dynamic exponent and ${\bf \bar q=q-Q_0}$ is the difference between the actual wave number ${\bf q}$ and ${\bf Q_0}$, the wave number where the susceptibility has a peak, $\chi_0({\bf Q_0},0)=max_q[\chi_0'({\bf q},0)]$.  If the OZ parameters are assumed to be $T$- and doping-independent\cite{23RSMMC,38Ande,39Naga,40Yama}, the resulting physics becomes quite simple.  For 2D materials, the Mermin-Wagner (MW) theorem\cite{41MerW} is satisfied, and the RPA transition at $T_{mf}$ turns into a pseudogap onset at $T^*\sim T_{mf}$, with a crossover to long-range order when interlayer coupling is strong enough -- in short, not much changes from the mean-field results.

When the OZ form of $\chi$ is assumed, $\chi_0(q)=\chi_0(Q)-A_2\bar q^2$, then $N_-$ becomes a constant, which we denote by $N_a$, and
\begin{equation}
\hat\lambda_2=\frac{a^2}{4\pi}\int \frac{d\bar q^2}{\delta +B\bar q^2} = N_a\int \frac{dX_-}{\delta +(X_--U_c)},
\label{eq:2d}
\end{equation} 
with $B=A_2U_c^2$.  Thus 
\begin{equation}
\xi_{th}^{-2}=\delta/B
\label{eq:2d2}
\end{equation} 
 and $N_a=a^2/4\pi B$.  
Eq.~\ref{eq:5} leads to long range order at $T=0$ only, with correlation length $\xi_{th}/a=1/\sqrt{4\pi N_a\delta}$ given by\cite{23RSMMC}
\begin{eqnarray}
\xi_{th} q_c=e^{T_2/T},
\label{eq:4}
\end{eqnarray}
where $T_2=\pi A_2\lambda /6\Gamma ua^2$, $A_i$ is the coefficient of $q^i$, and 
$q_c$ is a wave number cutoff.

The present $G_Z-W_Z-\Gamma_Z$ model is the simplest model which captures the essential physics of the pseudogap.  For a more quantitative comparison with experiment, two additional problems must be solved.  First, an extension to fully self-consistent $G-W-\Gamma$ may be needed to see how the susceptibility responds to the  gap-opening. Secondly, the term in Eq.~\ref{eq:3} proportional to $\omega^z$ must be included to describe quantum fluctuations.  Since we are primarily interested in the opening of the pseudogap at higher temperatures, we ignored it in our analysis.

\subsection*{Strong mode coupling leads to an extended range of short-range order}

Thus, mode coupling modifies the Stoner criterion to
\begin{eqnarray}
U\chi_0(q,0)=1+\lambda , 
\label{eq:5a}
\end{eqnarray}
where $\lambda$ satisfies Eq.~\ref{eq:5}.   In the OZ approximation, this becomes 
\begin{eqnarray}
\lambda \propto \int_{0}^{q_c} \frac{qdq}{q^2+\xi^{-2}} \sim ln(\xi q_c).
\label{eq:5b}
\end{eqnarray}
This OZ approximation misses the essential and strongly $T$-dependent mode-coupling physics, and we develop an alternative formulation of the theory.  In particular, the inverse curvature of the susceptibility $1/A_2$ can diverge due to a competition between conventional Fermi-surface nesting and Van Hove singularity (VHS) nesting, and this parameter plays a special role in the theory. In particular, we can get a semiquantitative picture of the correlation length changes by simply accounting for the $T$ dependence of $A_2(T)$.  We define the order-parameter DOS (OPDOS) as the threshold SDOS, 
\begin{eqnarray}
N_{OP}(T)=N_-(X_-=U_c^+)=a^2/4\pi A_2(T)U_c^2,
\label{eq:5c}
\end{eqnarray}
and use this in the OZ formalism, Eqs.~\ref{eq:2d}-\ref{eq:4}.  While this neglects the fact that $q_c$ can also have a strong $T$-dependence, and the contribution away from threshold is not negligible, the simplification should work when $\xi q_c>>1$.  We believe that this OPDOS quantifies the number of competing modes and provides an essential link to the SGA theory.  The connection between Eq.~\ref{eq:5} and entropy is further discussed in Supplementary Material Section III.

Figure~\ref{fig:3g} illustrates the profound effects that strong mode coupling has in LSCO, as well as the complete inability of the OZ approximation to capture this physics.  The SDOS, Fig.~\ref{fig:3g}(a), contains VHS-like features characteristic of conventional DOSs.  However, the singular behavior of Eq.~\ref{eq:5b} involves only features near the threshold, $X_-\sim U_c$, which evolve strongly with $T$, see inset to Fig.~\ref{fig:3g}(a).  For $T>0$, the threshold behavior is always a step at $X_-=U_c$, indicative of a parabolic peak in $\chi_0$ with curvature inversely proportional to the step height, Eq.~\ref{eq:5c}. For analyzing critical phenomena, we define a threshold correlation length $\xi_{th}$ (red solid lines in Figs.~\ref{fig:3g}(b)) in terms of the initial curvature of $\chi_0$ at ${\bf Q_0}$, Eq.~\ref{eq:2d2}.  We then evaluate $\xi_{th}$ using the full Eq.~\ref{eq:5}, but the results can be interpreted in terms of an OZ form with a strongly $T$-dependent step height $N_{OP}(T)$.   However, we caution that the experimentally measured $\xi_{1/2}$ is typically taken from the susceptibility half-width.  This $\xi_{1/2}$ is generally larger than $\xi_{th}$ due to the fast fall-off of $\chi$ near the edge of the $(\pi,\pi)$-plateau.  Thus, for the example of Fig.~\ref{fig:3g}(c), we calculated a $\xi_{1/2}$ (filled blue circles) from the full half-width of the renormalized $\chi$, and found that typically $\xi_{1/2}\sim 2\xi_{th}$, where the solid red line in Fig.~\ref{fig:3g}(c) plots $\xi_{th}$ vs $1/T$; the values of $\xi_{1/2}$ are in good agreement with experiment (green dot-dot-dashed line)\cite{42Birg}.  

\begin{figure}
\leavevmode
\rotatebox{0}{\scalebox{0.44}{\includegraphics{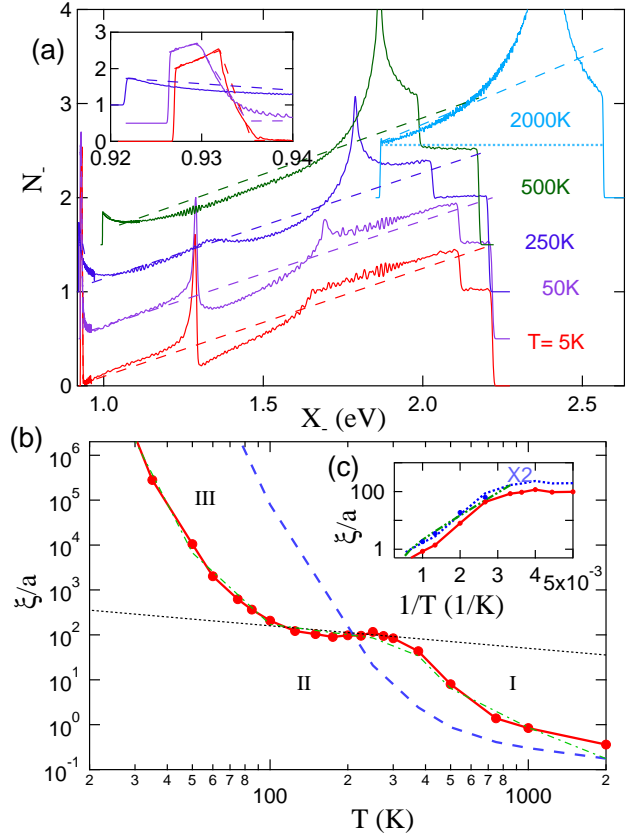}}}
\vskip0.5cm  
\caption{
{\bf Structures in LSCO correlation length.}  (a) SDOS for undoped LSCO at several temperatures; light-blue dotted line gives OZ form of SDOS.  (b) Corresponding temperature dependence of correlation length $\xi_{th}$. For $x=0$, the red solid line with filled dots is $\xi_{th}$ calculated from the solid lines in (a), while the thin green dot-dashed line is based on the dashed lines in (a), showing that $\xi_{th}$ is relatively insensitive to the structures away from threshold.  In contrast, the blue dashed line is based on the shape of the SDOS at $T=2000K$, but shifted and renormalized to match the SDOS at lower $T$, illustrating sensitivity to the leading edge structure.  
The black dotted line illustrates the scaling $\xi_{th}\propto T^{-1/2}$.
(c) Calculated $\xi_{th}$ replotted for $x=0$ (red solid line with filled circles) compared with $\xi_{1/2}$ (blue filled circles) and with experiment (green dot-dot-dashed line)\cite{42Birg}; the blue dotted line is twice $\xi_{th}$.  
} 
\label{fig:3g}
\end{figure}

Figure~\ref{fig:3g}(b) illustrates how strong mode coupling slows down the correlation length divergence.  Undoped LSCO (red solid line) shows two regions I and III of exponential growth of $\xi_{th}$ with decreasing $T$, separated by an anomalous region II where $\xi_{th}$ actually decreases with decreasing $T$.  
While the model captures the MW-like divergence at low $T$ (region III), our main interest is in the higher-$T$ behavior.  In the high-$T$ limit (region I), the leading-edge parabolic curvature is quite small, and if it were $T$-independent, as in the OZ approximation, the growth in $\xi_{th}$ would follow the blue dashed line (Eq.~\ref{eq:4}), but thermal broadening causes the curvature to decrease with increasing $T$, leading to the faster growth of the red solid line.  The origin of the anomalous region II for undoped LSCO, Fig.~\ref{fig:3g}(b), can be readily understood from Fig.~\ref{fig:3g}(a), where for $T<1500K$ there is excess SDOS weight near $U_c$, leading to a strong peak (inset) as $T\rightarrow 0$.  This feature represents the development of the $(\pi,\pi)$-plateau.   In Supplementary Material Section III we present a simple calculation of $\xi_{th}$, showing how a range of power-law behavior, $\xi_{th}\sim 1/T^{1/2}$ (dotted line in Fig.~\ref{fig:3g}(b)), could arise.  
While this power-law behavior captures the average $T$ dependence of the anomaly, it cannot reproduce the nonmonotonic variation of the correlation length.

\subsection*{Exploring parameter space}

The reference families can also be used to interpolate between physical cuprates.  Here we use interpolation to understand the origin of the condensation bottleneck found in region II.  We find that it is caused by {\it proximity} to a much stronger transition, a crossover between Mott and Slater physics, and that LSCO is more correlated than other cuprates, lying close to this crossover. Remarkably, this conclusion echoes an early neutron scattering study, which claimed that LSCO is close to an instability that is off of the physical parameter plane.\cite{43Aepp}

 We study two important cuts in $t'/t-t''/t$ space, a minimal cut ($t''=0$) and the Pavarini-Andersen [PA] cut ($t''=-t'/2$) -- the latter seems to best capture the physics of the cuprates.  By tuning $t'$ we unveil the origin of the condensation bottleneck as a localization-delocalization crossover tied to the crossover from $(\pi,\pi)$- to FS-nesting.  As a byproduct, we gain insight into why LSCO is so different from other cuprates, and how cuprates evolve from the pure Hubbard limit ($t'=0$).


\begin{figure}
\leavevmode
\rotatebox{0}{\scalebox{0.38}{\includegraphics{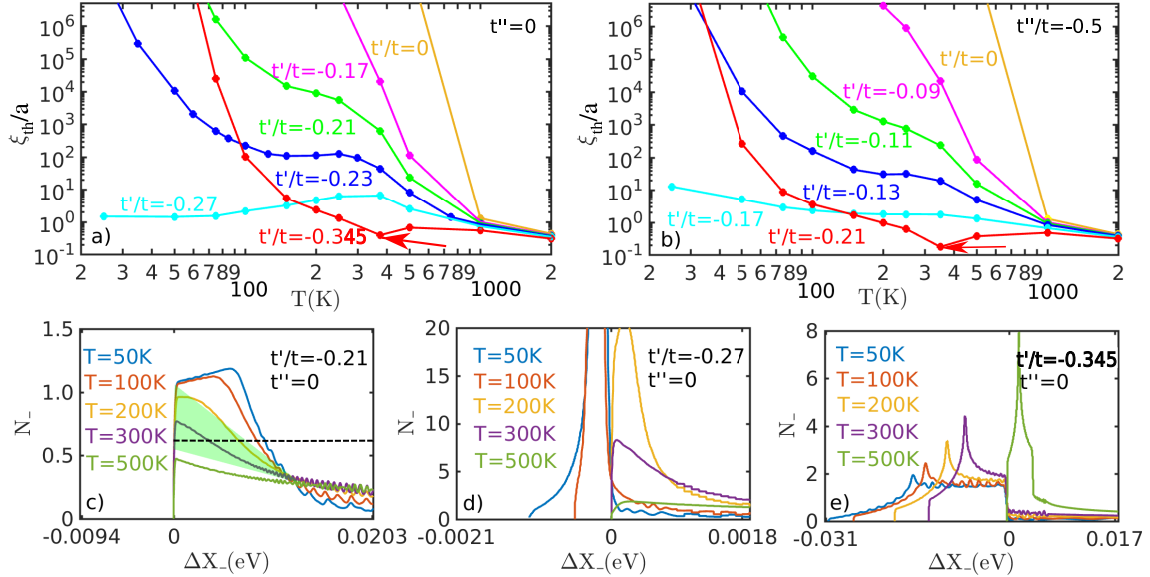}}}
\vskip0.5cm    
\caption{
{\bf t' dependence of correlation bottleneck.}  (a)   $\xi_{th}$ for the minimal reference family at several values of $t'/t$ [see legend] and $t''=0$.  (b) $\xi_{th}$ for the PA reference family at several values of $t'/t$ [see legend] and $t''=-t'/2$.  (c-e) SDOS at $t''=0$ and three values of $t'/t$ = -0.21 (b) [as in Fig.~\ref{fig:3g}(b)], -0.27 (c), and -0.35 (d).  All curves are shifted to line up the $(\pi,\pi)$ data at $\Delta X_- = X_- - X_-(\pi,\pi) =0$.  Arrows in (a) and (b) show dip in $\xi_{th}$ associated with the commensurate-incommensurate transition. 
} 
\label{fig:3g2}
\end{figure}


Figure~\ref{fig:3g2}(a) shows the $T$-evolution of $\xi_{th}$ for several values of $t'$ along the minimal cut, including the data of Fig.~\ref{fig:3g}(b).  For $t'/t >-0.17$, the system is characterized by commensurate $(\pi,\pi)$ order with an effective Neel temperature $T_N\sim$1000K [the correlation length grows so rapidly that interlayer correlations will drive a transition to full 3D order].  Similarly, for $t'/t<-0.345$ there is incommensurate $(\pi,\pi-\delta)$ order with $T_N$ about a factor of 10 smaller.  But for intermediate $t'/t$ the {\it commensurate-incommensurate transition} is highly anomalous, with correlation length orders of magnitude smaller than expected.  Figure~\ref{fig:3g2}(b) shows that a similar evolution follows along the PA cut in parameter space.  The reason for this anomalous behavior can be seen by looking at the leading edge SDOS in the crossover regime, shown for three values of $t'$ along the minimal cut in Figs.~\ref{fig:3g2}(c)-(e).  For ease in viewing, these curves have been shifted to line up the SDOS at $(\pi,\pi)$ at all $T$.  It is seen that the anomalous collapse of $\xi_{th}$ is associated with a rapid growth of the step height $N_{OP}\sim 1/A_2$, culminating in a near-divergence at $t'_c=-0.27t$, where the leading edge curvature $A_2$ goes to zero, Fig.~\ref{fig:3g2N}(b).  Note that $\xi_{th}$ drops by 9 orders of magnitude at $T=200K$ when $t'/t$ changes from -0.17 to -0.27, then grows by a similar amount at 100K when $t'/t$ changes from -0.27 to -0.345. This transition may be hard to detect experimentally, since over this same range, the $\xi_{1/2}$ will be frozen at the value corresponding to the half-width of the $(\pi,\pi)$ plateau.  The green shaded region in Fig.~\ref{fig:3g2}(c) shows that the range of the anomalous growth (II) of $\xi_{1/2}$ in Fig.~\ref{fig:3g}(b) coincides with the range of rapid growth of the SDOS leading edge; such behavior is absent if a $T$-independent OZ form (black dashed line) is assumed.  

From the relationship between $\xi_{th}$ and the step height $A_2$, Eq.~\ref{eq:4}, we see that a diverging $A_2$ will cause $\xi_{th}\rightarrow 0$.  The divergence arises when the susceptibility at $(\pi,\pi)$ crosses over from a maximum to a local minimum.  In the latter case, the maximum intensity of $\chi (q,0)$ is spread along a `ring' in $q$-space surrounding $(\pi,\pi)$, so that the inverse susceptibility resembles a `Mexican hat',  Fig.~\ref{fig:3g2N}(c).  Approximating the $\chi^{-1}$-dispersion by a Mexican hat form $\chi^{-1}_0-A_2q^2+A_4q^4$ leads to a threshold divergence $N_-\sim (\chi^{-1}-U_c)^{-1/2}$ as $A_2$ passes through zero, similar to Fig.~\ref{fig:3g2}(d) or Fig.~\ref{fig:3b}(a).  Note that the SDOS-divergence resembles a 1D VHS in the conventional DOS, even though here the 1-D direction is the radial direction away from $(\pi,\pi)$. The divergence of $\chi$ along a ring can be thought of as a 2D analog of Overhauser's effect\cite{24Over}.   Similar effects are found in Bi2201, next Subsection.

The strength of the anomaly in $\xi_{th}$ at $t'_{c}=-0.27t$ suggests that it is not an ordinary quantum critical point.  Indeed, from Fig.~\ref{fig:009c} it can be seen to be a `supertransition' between two domains of attraction, one commensurate at $(\pi,\pi)$, the other incommensurate at $(\pi.\pi-\delta)$ (compare Figs.~\ref{fig:009c}(a) and (b) at $x=0$).  Such a supertransition is known from spectrum-generating algebras (SGAs) to be highly anomalous.  The diverging OPDOS at the supertransition, also characteristic of SGAs, indicates that many different $q$-vectors compete simultaneously, frustrating the divergence of any particular mode.  This can be clearly seen in Fig.~\ref{fig:3g2N}(b), where the susceptibility is nearly flat.  This is the electronic analog of McMillan's phonon entropy: if many phonons are simultaneously excited, the transition is suppressed to anomalously low temperatures.  At this transition the {\it ground state is infinitely degenerate}, representing an anomalous spin frustrated state arising in the absence of disorder, with the frustration arising from strong mode coupling, although whether it represents a spin liquid or spin glass requires further analysis.  
  Note also that when the commensurate-incommensurate transition is at $T>~300K$, $\xi_{th}(T)$ has a sharp downward cusp at the transition, while above the transition $\xi_{th}$ is strongly suppressed, $\xi_{th} < a$, arrows in Figs.~\ref{fig:3g2}(a),~(b).


 For LSCO, the anomaly in Fig.~\ref{fig:3g}(b) (large red dot in Fig.~\ref{fig:2c}(c)) seems to fall near the coherent-incoherent crossover (pink shaded region), consistent with an early neutron scattering result which indicated that doped LSCO is close to a magnetic QCP.\cite{43Aepp}  This proximity to a novel disorder-free spin-glass or spin-liquid QCP may play a role in stripe physics.  Notably, in LSCO, commensurate $(\pi,\pi)$ order disappears rapidly by $\sim 2\%$ doping, being replaced by incommensurate magnetic fluctuations and low-$T$ spin-glass effects.   Moreover, the commensurate-incommensurate transition involves a highly disordered regime separating two well-ordered phases (Figs.~\ref{fig:3g2}(a)~,(b)).  Since ordering tends to lower the free energy of the electronic system, the disordered regime represents a state of high free energy, and doping across this region can lead to a regime of [nanoscale] phase separation (NPS), which in LSCO is manifest as the stripe phase.\cite{44MistNPS}  It should be noted that also in Cr the commensurate-incommensurate AF transition is first order\cite{Fawcett}.  The very different situation in most other cuprates is discussed in the next Subsection, where it is shown that high-$T$ magnetic fluctuations give way at low $T$ to a fragile CDW order, consistent with experiment.  

Finally, we note that the $t'$ of the Mott-Slater transition  is also the point at which the DOS at the bottom of the band changes from a local minimum to a local maximum (green dotted arrow in the inset to Fig.\ref{fig:009d}(c) -- see also Fig.~\ref{fig:3g2N}), and close to the point at which a nematic transition [splitting the degeneracy of the VHSs at $(0,\pi)$ and $(\pi,0)$] becomes possible.\cite{Gonzalez}

 \subsection*{Mode coupling in Bi2201}


We noted earlier that the OZ form should not be used in quantum critical theory when FS nesting is present\cite{36Mill}.  The problem can however be addressed by using SDOS.  The issue is that at $T=0$ $\chi_0$ is nonanalytic, with cusp-like peaks, Fig.~\ref{fig:2b}, so the leading $q$-dependence is not parabolic.  Indeed, since the dominant instabilities tend to be associated with double-nesting, the peak shape can be quite anisotropic, $\sim |\bar q|$ in some directions and $\sim |\bar q|^{1/2}$ in others\cite{Gzm1}.  There is a simple relation between the $\bar q$-dependence of the peak susceptibility and the form of $N_-(X_-)$ at threshold.  If the SDOS varies as a power law, $N_-={\partial\bar q^2/\partial X_-}\sim X_-^p$, then at threshold, $(X_--U_c)\sim\bar q^{2/(p+1)}+\xi_{th}^{-2/(p+1)}$.  For a parabolic onset $(X_--U_c)\sim\bar q^2+\xi_{th}^{-2}$, or a step in $N_-$ at threshold ($p=0$ or $N_-=const>0$ for $X_->U_c$).  Instead, for Bi2201 there are cusps in $\chi_0$ near $(\pi,\pi-\delta)$ and an antinodal cusp at $(\delta,\delta)$. For both cusps, a smooth power law is found, with $p\sim 2$, or $(X_--U_c)\sim |\bar q|^{2/3}+\xi_{th}^{-2/3}$ at threshold.

However, this leads to new problems.  When a term $|\bar q^p|$ for $p\le 1$ is substituted into Eq.~\ref{eq:5}, the integral converges at threshold, so the Mermin-Wagner effect appears to be absent.  That this is not the case can be seen from Fig.~\ref{fig:2b}: at any finite $T$, the cusp is rounded off, and $\chi_0$ varies parabolically, which means $A_2$ is finite.  The cusps still produce a strong effect in that $A_2$ has an exceptionally large $T$-dependence, $A_2\rightarrow\infty$ as $T\rightarrow 0$.  This leads to an extra strong divergence of $\xi_{th}$, Eq.~\ref{eq:4}, at $T\rightarrow 0$.  But, as we increase $T$ from zero, this also means that the corresponding $\xi_{th}$ decreases rapidly, leading to a very fragile phase.  We have separately confirmed that the order is also sensitive to $\omega$ and disorder.

\begin{figure}
\leavevmode
\rotatebox{0}{\scalebox{0.44}{\includegraphics{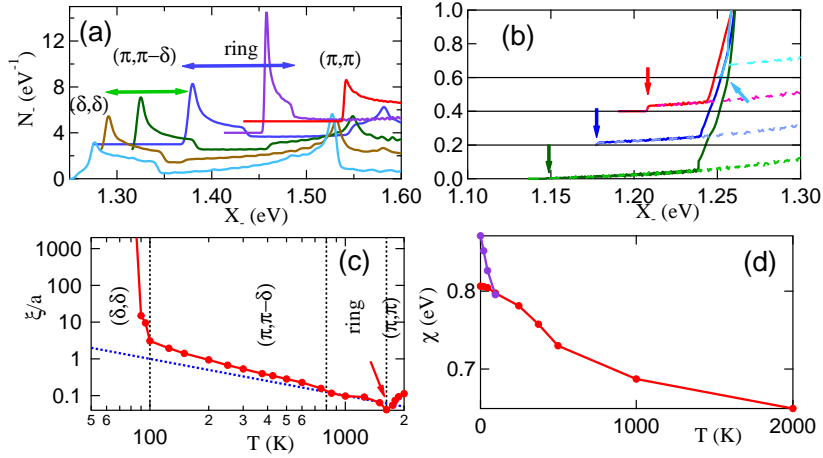}}}
\vskip0.5cm  
\caption{
{\bf Structures in Bi2201 correlation lengths.}
Temperature dependence of: (a) SDOS $N_{-}$ for $t-t'-t''$ reference state of Bi2201, $x=0.20$. Datasets are offset for clarity; from highest to lowest, $T$ = 2000 (red), 1000 (violet), 500 (blue), 375 (green), 250 (brown), and 100K (light blue).  (b) Continuation of SDOS in (a) to lower $T$, showing evolution of the leading edge.  From top to bottom: $T$ = 100 (light blue), 50 (red), 25 (blue), and 5K (green).  For each curve, a partial SDOS near the ANN peak is superposed to show the SDOS associated with the ANN peak whose onset is marked by arrows.  (c) Correlation length vs $T$.  Blue dotted line represents $\xi_{1/2}\sim T^{-1}$. Arrow shows dip in $\xi_{1/2}$ associated with commensurate-incommensurate transition. (d) Susceptibility of $\chi (q)$ peak vs $T$ for $q$ near the $(\pi,\pi)$-plateau (red line) or for ANN nesting (blue line).
} 
\label{fig:3b}
\end{figure}

In Bi2201 and most other cuprates, a larger ratio $|t'/t|$ expands the $(\pi,\pi)$-plateau, leading to a rich phase diagram, Fig.~\ref{fig:009c}(b).  In this case the b-VHS is dominant only at high $T$, while FS-related susceptibility features cause a crossover of the $(\pi,\pi)$ susceptibility from a maximum at high $T$ to a local minimum at lower $T$.  This leads to a different kind of strong mode coupling at intermediate $T$, best exemplified by the ring phase, Figs.~\ref{fig:009}(d,f).   In this phase, the leading-edge SDOS has a `Mexican hat'-like divergence (previous Subsection), Fig.~\ref{fig:3b}(a) near 2000K, leading to a slow growth in $\xi_{th}\sim T^{-1}$, Fig.~\ref{fig:3b}(c), quite similar to that observed in doped YBCO\cite{Ouazi,Alloul}. Near the onset of ring order, $T=1625K$, the correlation length has a nonmonotonic $T$-dependence, similar to that seen in LSCO, Fig.~\ref{fig:3g2}(a) and (b).

While many pseudogap-like features are similar to LSCO, Bi2201 has some novel properties associated with the ANN instability, which can become a CDW when charge order is included in the model.  Since the ANN susceptibility peak has a much larger intrinsic curvature, mode-coupling effects are weak.  Hence the ANN susceptibility peak grows rapidly at low temperature, and as doping increases, it can surpass the $(\pi,\pi)$-peak, thereby sharply cutting off the growth of near-$(\pi,\pi)$ fluctuations. This is illustrated in Fig.~\ref{fig:3b} for $x$ = 0.20.   Notably, since this peak is not associated with strong mode coupling it has a weak SDOS with a very strong $T$-dependence, Figs.~\ref{fig:3b}(b) and~\ref{fig:3b}(d), crossing the AFM peak near $T=100$K.  The large subsequent growth of the ANN correlation length would ensure the rapid establishment of a true long-range 3D ANN order.  However, corresponding to the large $T$-dependence of $\chi_{ANN}$, there is a similar strong disorder dependence -- particularly in the presence of short-range near-$(\pi,\pi)$ order, which should cause the ANN correlation length to saturate.

\section*{Discussion}

\subsection*{Cuprate pseudogap}

In Figs.~\ref{fig:2cN} and~\ref{fig:2c}, and Results Section~(DFT-Lindhard), we discussed a revised VHS scenario in which a number of VHS related features evolve nearly linearly in doping and scale with $T_{VHS}$. We noted that this scaling resembles the pseudogap crossover $T^*(x)$.  Here we provide additional evidence for this identification.  Note that $T_{VHS}$ is typically close to the measured $T^*$ in most cuprates, including YBCO and Hg-cuprates\cite{9Doir} (yellow-green  line in Fig.~\ref{fig:2c}) and LSCO (magenta dotted line).\cite{Tall} Furthermore, the pseudogaps in Bi2201 and Bi2212 terminate near the point $T_{VHS}\rightarrow 0$, i.e., at the conventional VHS\cite{45Kusk,46DasOp,47Piri,48Niem,49Benh}.

The phase diagram of Fig.~\ref{fig:2c} has a number of further consequences for cuprate physics.  First, the VHS onset near $T_{VHS}$ combined with the coherent - incoherent crossover (pink shaded region in Fig.~\ref{fig:2c}) can explain the anomalous transport properties found near the pseudogap, Fig.~\ref{fig:2cN}. This parallels transport, where for $T>T^*$, the resistivity $\rho$ varies linearly with $T$,\cite{9Doir}, a  behavior expected near a VHS\cite{50Buhm}.  For lower $T$, the resistivity is mixed, but $\rho\sim T^2$, as expected for a coherent Fermi liquid below a $T_{coh}<T_{VHS}$.  The brown long-dashed  line in Fig.~\ref{fig:2c} represents the corresponding $T_{coh}$ found in YBCO and Hg-cuprates\cite{9Doir}, in excellent agreement with the coherent-incoherent crossover.  

We note further that the compressibility should have a peak similar to $\gamma$ that extrapolates to the VHS doping as $T\rightarrow 0$, consistent with our Fig.~\ref{fig:2c}.  In a related cluster-DFT calculation\cite{Tremblay}, the pseudogap is associated with a Widom line, related to a peak in the compressibility. This is further discussed in Supplementary Material Section IV.
Notably, the Widom line terminates in a first-order phase transition near the VHS. The similarity of this phase diagram to the excitonic liquid-gas transition found in semiconductors\cite{JPW} suggests that the b-VHS may be best described as a VHS-exciton\cite{ExIns1,ExIns2}.

This picture bears a resemblance to the Barzykin-Pines model of the cuprate pseudogap,\cite{51Barz} identifying $T_{VHS}$ and the coherent - incoherent crossover with $T^*$ and $T_{coh}\sim T^*/3$ in their model.  Since their model is related to Kondo lattice physics\cite{52Curr}, the  question arises as to whether a similar mode-coupling calculation in heavy-fermion compounds could lead to a similar anomalous entanglement at the f-electron incoherent-to-coherent transition.

\subsection*{Classifying order parameters}

Competition between order parameters also arises in a totally different field, nuclear physics, where order parameters are formally classified in terms of a spectrum generating algebra (SGA) (Supplementary Material Section I.C). Order parameters are grouped into families (Lie subalgebras), and an order parameter density-of-states (OPDOS) is defined.  In addition to conventional phase transitions, there are `supertransitions' (called excited-state quantum critical points) where the system is driven across a boundary between two families and the OPDOS diverges.  As a byproduct of this work, we develop a similar classification of order parameters for condensed matter systems, where similar phenomena can be observed (Supplementary Material Section I.A).

\subsection*{Strong coupling physics}

The condensation bottleneck that we identified in Fig.~\ref{fig:3g2} appears to be a previously unexpected {\it transition between Mott and Slater physics}, shedding light on the controversies between Anderson and Scalapino.\cite{54And,55Scal}  On the Slater-side of the line (larger $|t'|$), the cuprate physics is dominated by FS nesting with Neel temperatures (strictly, $\xi_{th}/a\sim 1000$) near 100K, while on the Mott side, the Fermi surface plays no role, and the magnetic transition is at $T\sim$~1000K, and always lies at $(\pi,\pi)$ due to VHS nesting.  Thus, the phenomenological strong-coupling Yang-Rice-Zhang model\cite{56YRZ} can only work on the Mott side of the transition, since it requires $(\pi,\pi)$ nesting.  However, at the crossover, the correlation length remains nanoscale, leading to strong mixing of the two competing phases.  Among the physical cuprates, all fall into the Slater domain, except for LSCO, which lies close to the Mott phase.  Notably, LSCO shows the strongest signs of nanoscale phase separation in the form of stripe physics [Results Section (Beyond RVB)].  This result is consistent with a recent first-principles calculation of the Hubbard $U$, which finds LSCO more strongly correlated than other cuprates.\cite{57Jang}

We recall that our self-energy formalism\cite{AIP} is able to reproduce most spectral features of the cuprates in terms of competition between d-wave superconductivity, a $(\pi,\pi)$ ordered AFM phase, and an ANN CDW.  We reproduce not only the photoemission dispersions, limited to the lower Hubbard bands, but optical and x-ray spectra, which depend sensitively on the Mott gap.  
In a related 3-band model, we reproduced the Zhang-Rice result that the first doped holes are predominantly of oxygen character, with overall dispersions at half-filling agreeing with subsequent DMFT results\cite{AIP}. 
While our earlier calculations predicted long-range $(\pi,\pi)$ AFM order to persist to too high $T$, this is corrected in the present calculations which have only short range order and reproduce the upper and lower Hubbard band dispersions with broadening $\sim 1/\xi_{th}$.\cite{23RSMMC}
Moreover, the calculated $\xi_{1/2}$ values (filled blue dots in Fig.~\ref{fig:3g}(c)) agree with experiment in LSCO (green dot-dot-dashed line)\cite{42Birg}.  The experimental data are shown only above 300K, since at lower $T$ interlayer coupling drives a transition to long-range order.  In the Heisenberg model, $\xi_{1/2}$ is a function only of $T/J$, and these data have been used to adduce the value of the exchange $J$.  Hence, our mode coupling calculation successfully reproduces the crossover from the $U$-scale to the $J$-scale physics, as was found earlier for electron-doped cuprates\cite{23RSMMC}.

For more insight into the strong coupling limit, we note that in the two-particle self-consistent approach\cite{58Vilk}, $\lambda$ is determined by a sum rule involving double occupancy, which is fixed by assuming 
$$U_{sp}/U=<n_{\uparrow}n_{\downarrow}>/(n/2)^2,$$
which leads to a saturation of $U_{sp}$ as $U\rightarrow\infty$ (or $<n_{\uparrow}n_{\downarrow}>\sim 1/U\sim J$).  In our calculation, this saturation arises naturally, since $U_{sp}$ can never exceed $U_c$.

\subsection*{VHS and excitons}

While we have demonstrated the important role of the VHS in pseudogap physics, we have not addressed the deeper question of the physical significance of the VHS.  We suggest that the significance is primarily topological: the saddle-point VHS is associated with the crossover of a Fermi surface from electron-like to hole-like, and hence represents the point nearest to electron-hole symmetry.  This explains its strong role in transport, Fig.~\ref{fig:2cN}.  While electron-electron scattering is strong, it usually plays only a small role in transport, since if an electron-electron scattering conserves energy and momentum, it does not change the net current.  In contrast, any electron-hole collision has a large effect on current.\cite{Baber}  

Now strong electron-hole scattering is indicative of a tendency to exciton formation, and we believe our results can most easily be understood in terms of a VHS-driven excitonic instability.  Thus, the large entropic effects are related to a broad spread of the electrons in $k$-space, consistent with a localization of the real-space electron-hole separation in an exciton.  A two-band excitonic insulator transition is analogous to a superconducting BCS-BEC transition; we suggest that in our one-band case this becomes the Mott-Slater transition we have found, with the Mott (BEC) phase dominated by preformed excitons.  It is often stated that excitonic effects are unimportant in metals due to screening.  However, we are studying the Hubbard model, where excitons could be present near half-filling, and their evolution with doping could contribute to a metal-insulator transition.

\subsection*{Conclusions}

The present paper has provided significant progress toward DFT corrected MBPT.  (1) We have demonstrated the significant role of bosonic entropy in driving short-range order. Our computational scheme incorporates the DFT-Lindhard function, and should capture strong correlation effects, at least for $T_N\sim 1/U$.  (2) We have shown that the DFT-Lindhard function can be used to classify phase transitions, and may yield a condensed matter analog of SGAs.  (3) We have shown that entropy effects in a one-band model involve a non-FS contribution associated with the VHS.  In turn, these entropic effects modify the VHS physics, causing the spectral weight associated with the VHS [in the heat capacity and susceptibility] to shift rapidly towards half-filling as $T$ increases, instead of following the evolution of the ($q=0$) DOS.  This large shift suggests that the role of the VHS in pseudogap physics needs to be reexamined.  

Finally, we comment on the limitations of the present mode-coupling framework. In particular, one would like to go beyond the present results to full inclusion of vertex corrections and {\it excitonic physics}.    
Another issue is the {\it evolution of the Fermi surface in the regime of short-range order}.  As the FS shrinks from a large to a small size, the susceptibility will also change, and drive changes in the fluctuation maps and the associated properties discussed in this study such as the appearance of a second phase out of a regime of short-range order of a primary phase.  With a DFT-corrected Lindhard function, one should be able to obtain insight into these issues.


\section*{Acknowledgements}     
This work is supported by the US Department of Energy, Office of Science, Basic Energy Sciences grant number DE-FG02-07ER46352, and benefited from Northeastern University's Advanced Scientific Computation Center (ASCC) 
and the allocation of supercomputer time at NERSC through grant number DE-AC02-05CH11231.  Part of this work was done while RSM was on sabbatical at the Advanced Light Source in Berkeley and the Los Alamos National Laboratory, where he benefitted from numerous discussions.  

\section*{Author contributions}
R.S.M., I.G.B., P.M., C.L., and A.B. contributed to the research reported in this study and the writing of the manuscript.

\section*{Additional information}
The authors declare no competing financial interests. 
\end{document}


\title{{\color{red}SUPPLEMENTARY MATERIAL for}\\
Entropic Origin of Pseudogap Physics and a Mott-Slater Transition in Cuprates}
\author{R.S. Markiewicz}
\author{I.G. Buda}
\author{P. Mistark}
\author{C. Lane}
\author{A. Bansil}

\maketitle


\section{Order parameter classification scheme}

\subsection{Classifying order parameters}

Our order parameter classification scheme is based on the existence of a generalized Stoner criterion for phase instability, 
\begin{equation}
V_q\chi_0(q,\omega=0)=1+\lambda. 
\label{eq:A1}
\end{equation}
The restriction to zero frequency means that we cannot discuss quantum critical phenomena; we shall see that the formulation works best at higher $T$s.  In analyzing Eq.~\ref{eq:A1}, we assume that the $q$ dependence of $V_q$ is weak.  This is certainly true in the Hubbard model.  For the bare Coulomb interaction $V_q$ behaves like a power law, e.g., $V_q\sim 4\pi e^2/q^2$ in 3D, and the only special $q$ values are 0 and $\infty$.  We can then calculate {\it fluctuation maps} that divide the $T-x$ or the $\omega-x$ phase space into {\it basins of attraction}, where the $q$-vector corresponding to the maximum $\chi_0(q,0)$ value changes character.  Examples of fluctuation maps are shown in Figs.~3 and~4 of the Main text.  Phase transitions have a different character depending on whether they arise within a basin or are due to transitions across the boundary  between different domains.  In spectrum generating algebras, this boundary is referred to as a {\it separatrix}.\cite{SGA1,SGA2}

For phase transitions within a single basin, the transition is controlled by the net electron degeneracy, and hence typically by an SO(N) point group, and by the corresponding $U$s that break the degeneracy.  For example, if we consider density wave fluctuations at a particular $Q$, the 1 on the right hand side of Eq.~\ref{eq:A1} is appropriate for a SDW, but shoud be -1 for a CDW.  The difference is an issue of charge neutrality.  The Coulomb energy is so large that if electrons bunch up into a CDW it will cost an energy comparable to the plasma energy, unless $V$ is attractive.  On the other hand, in an SDW the up and down spins bunch up out of phase, so the magnetization oscillates in space while the charge density is constant, and hence condensation is easier.  The same effect holds if there is an orbital or valley degeneracy.  Moreover, in a Peierls-like scenario, the electron bunching is accompanied by a lattice distortion controlled by a Stoner condition equivalent to Eq.~\ref{eq:A1}, with an effective interaction $U_{eff}$.  For the cuprates, the two VHSs at $(\pi,0)$ and $(0,\pi)$ act as a form of valley degeneracy, and the theory with $q=(\pi,\pi)$ is controlled by an SO(8) group.  Similarly, for antinodal nesting the $(\delta,\delta)$ density wave can be either SDW or CDW.

A new result of our investigation is the possibility of {\it supertransitions}, where doping or temperature causes the system to cross a separatrix between different basins of attraction.  In the theory of spectrum generating algebras these are known as {\it finite energy quantum critical points}, and it is found that an {\it order-parameter DOS} diverges at the transition.  We find that we can reproduce similar behavior by defining the order-parameter DOS as the SDOS at threshold, which controls the singular behavior of $\xi$ [Results Section (Beyond RVB) of the Main text].

Finally, we discuss how the fluctuation maps can be used to generate phase diagrams.  At mean field level, a phase transition occurs when the generalized Stoner criterion of Eq.~\ref{eq:A1} is satisfied, so that if $V_q$ is a constant, the transition would fall along one of the contour lines (black dots) in Fig.~2 of the Main text, and even if $V_q$ varies with $x$ and/or $T$, these lines can be used to find the transition.   For a 3D system, this will simply mean that mode coupling depresses the transition temperature to a lower $T$ where the Stoner criterion is satisfied, leaving behind an extended pseudogap regime.  Once the system develops long-range order, the Fermi surface and susceptibility change, and the fluctuation map at lower $T$ must be recalculated.  Does this mean that transitions predicted from the fluctuation maps are inaccessible?  If the lower-$T$ phase actually has the lower free energy, a transition might occur spontaneously, or the system might get trapped in a metastable state.  In this case, there are two possible ways of observing these states.  First, since there is an equivalence class of materials with the same fluctuation map, it is possible that different members of the class have smaller values of $V_q$, for which the Stoner criterion is only satisfied in the lower-$T$ phase.  Alternatively, one might try to quench or photoexcite the system.

For a 2D system, the issue is more subtle as the Mermin-Wagner theorem holds and an ordered phase can only arise at $T=0$ unless the system transitions to 3D or its spin dimension decreases due to spin-orbit coupling. 
Thus, when the correlations of one phase start to grow, one will need to account for the evolution of the Fermi surface and susceptibility at lower $T$. This can be complicated as each competing density wave corresponds to a different Fermi surface reconstruction.  Here we note one point in particular.  In the Main text, Fig.~15(c), we see the development of an ANN phase out of a short-range order $(\pi,\pi-\delta)$ SDW in good agreement with the CDW phase found in many cuprates.   Since the SDW correlation length is only $\xi_{SDW}\sim 5a$, this is perhaps not too surprising, but for lower $x$ $\xi_{SDW}$ should grow, and the developing small Fermi surface should be taken into account.  One issue is whether the CDW develops homogeneously or on the SDW domain walls.  In this connection we note some experimental evidence of a possible relevant phase found in another cuprate.  In LSCO, the low-temperature tetragonal (LTT) phase is believed to be highly relevant to stripe physics, anchoring the direction of the stripe to the lattice and producing the unusual criss-cross stripe pattern along the c-axis.  Yet the LTT phase is almost never the stable phase, existing as fluctuations or as a minority phase, although it can be stabilized via rare-earth substitution.  Electron microscopy studies find that the LTT phase tends to form as inclusions or on twin-boundaries of the low-temperature orthorhombic phase\cite{LTT1,LTT2}, suggesting that quenching these materials into the low-$T$ phase might prove interesting.

Lastly, one can apply tuning parameters to selectively turn an order on or off.  We have seen that disorder can be used to quench the ANN phase.  Moreover, magnetic fields are regularly used to turn off superconductivity to reveal the competing phases.  
A magnetic field has an inordinately strong effect in destroying spin-singlet Cooper pairs.  For instance, at high fields, superconductivity is often restricted by the Clogston-Chandrasekhar limit\cite{Clogston,Chandrasekhar}, $\mu_BH=\Delta(0)/\sqrt(2) \sim 1.76k_BT_c/\sqrt(2)\sim$ 2.6meV, assuming $T_c =30$K, superconducting gap $\Delta(T)$, and the BCS ratio for $2\Delta(T=0)/k_BT_c$.  Since the Bohr magneton is $\mu_B=0.057$meV/T, this gives a critical field ~46T, in good agreement with experiment in LSCO\cite{Taillefer}. In contrast, magnetic fields have relatively weak effects on antiferromagnetism\cite{Fawcett}.  Typically, one expects a spin-flop transition at fairly low fields, where the AF moment aligns perpendicular to $H$, followed by a gradual rotation of the moments until they are parallel to B and the AF order is destroyed.  While a weak spin-flop transition is often observed in the undoped cuprates, associated with Dzyaloshinskii-Moriya induced canting of the moments, the main effect of the field is only expected near the saturation field, $\mu_BH \sim J\sim$ 100meV $\sim$1800T.

\subsection{$x-\omega$-scaling}

\begin{figure}
\leavevmode  
\epsfig{file=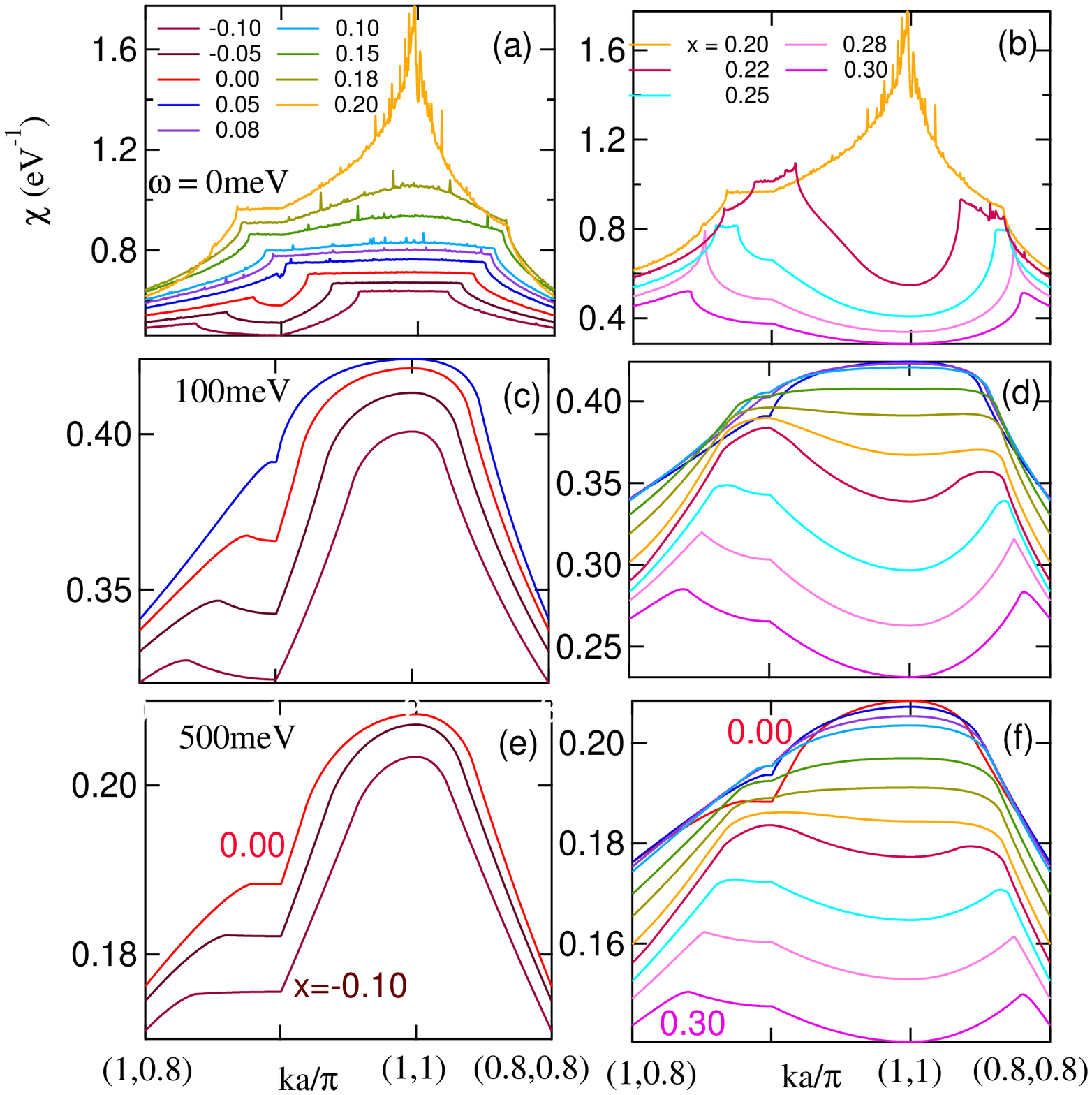,width=9.5cm, angle=0}
\vskip0.5cm
\caption{{\bf Doping dependence of $\chi_0$ at fixed $\omega$ and $T=0$.} Left hand frames are for $x<x_{peak}$ and right hand frames for $x>x_{peak}$. Value of $x$ is color coded according to legends in frames (a) and (b), and $x_{peak}$ is plotted in Fig.~\ref{fig:6}. 
Frequency $\omega$ = 0 (a,b), 100~meV (c,d), and 500~meV (e,f).   Calculations are based on full LDA-based dispersion for LSCO.
} 
\label{fig:4} 
\end{figure}
\begin{figure}
\leavevmode  
\epsfig{file=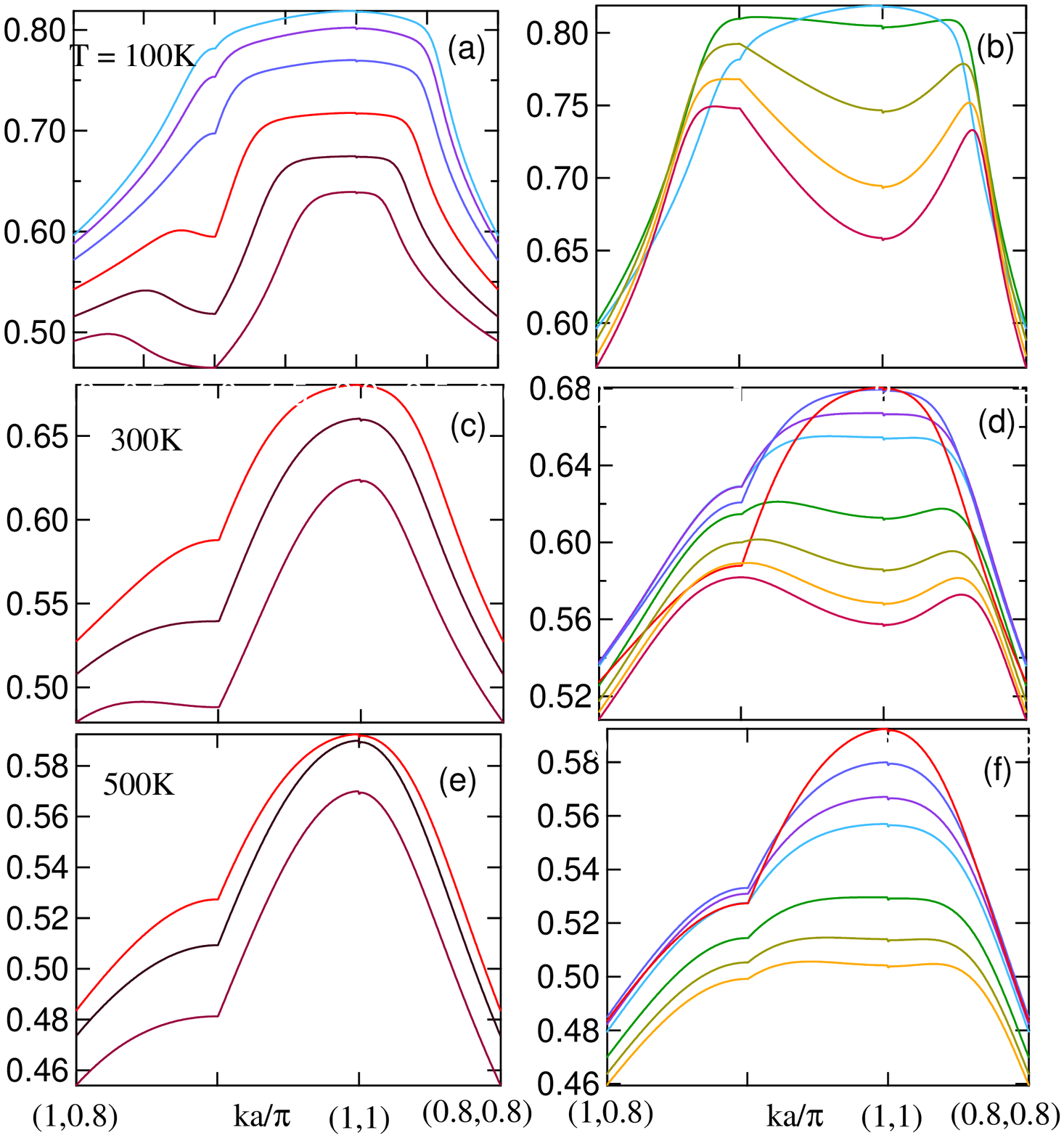,width=9.5cm, angle=0}
\vskip0.5cm
\caption{{\bf Doping dependence of $\chi_0$ for fixed $T$ at $\omega=0$.} Setup and doping color scheme is the same as for Fig.~\ref{fig:4}, except temperature $T$ =  100K (a,b), 300K (c,d), and 500K (e,f).   
} 
\label{fig:5} 
\end{figure}
\begin{figure}
\leavevmode  
\epsfig{file=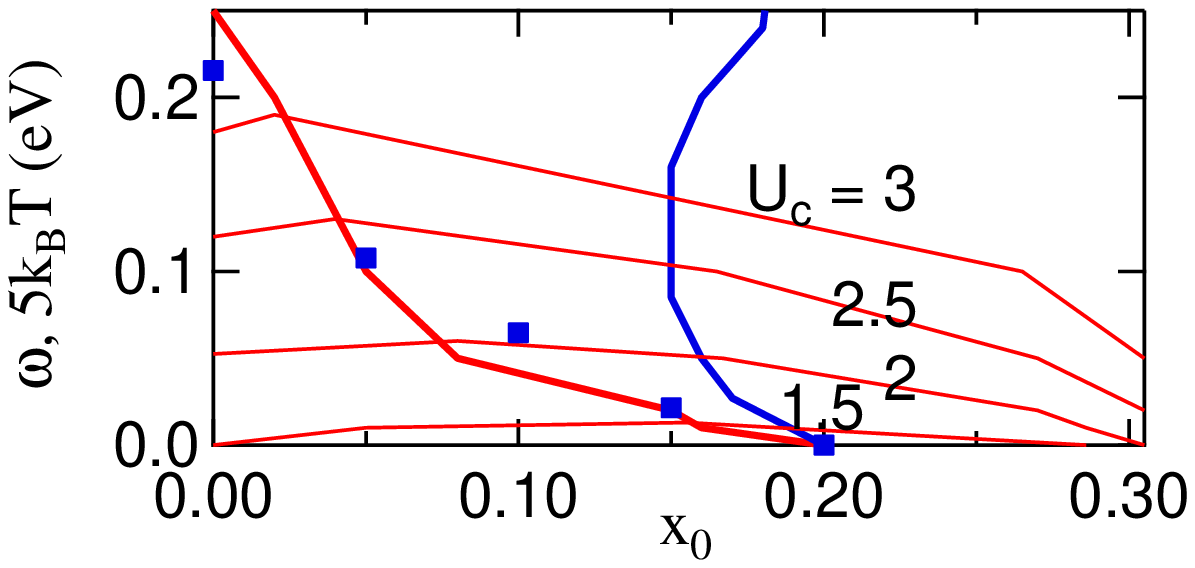,width=9.5cm, angle=0}
\vskip0.5cm
\caption{{\bf Commensurate-incommensurate crossover in the $x-\omega$ space.}
Plot of $x_{peak}$ as a function of $\omega$, taken from Fig.~\ref{fig:4} and related calculations (red solid line), compared to the scaled results from the Main text, Fig.~9 [blue squares], and of the commensurate-incommensurate transition (blue solid line).  Thin red lines indicate contours of constant $\chi_0$, labeled by their equivalent values of $U_c=1/\chi_0$. All curves are plotted vs frequency $\omega$, except the blue squares, which are plotted vs 5$k_BT$.
} 
\label{fig:6} 
\end{figure}

It is important to note that for a Fermi liquid, the scaling in the $x-T$ space found in the Main text (Fig.~9) implies a similar scaling in the $x-\omega$ space.  In fermionic systems there is a similar scaling in $T$ and $\omega$, with scale-factor 
\begin{equation}
\hbar\omega\sim ak_BT,
\label{eq:F1}
\end{equation}$a\sim\pi$.  Here we compare the $x-T$ map of Fig.~M9 with the corresponding $x-\omega$ maps, which can be more readily compared with the SGA results.      First, in Figs.~\ref{fig:4} and~\ref{fig:5} we compare plots of susceptibility curves varying $x$ at either $T=0$ and fixed $\omega$ (Fig.~\ref{fig:4}), or at $\omega = 0$ for several values of $T$ (Fig.~\ref{fig:5}).  The latter were used to calculate the light-blue dot-dashed line in the Main text, Fig.~9. For each $\omega$, starting at electron doping $(x<0)$, there is a peak at $(\pi,\pi)$ which grows with $x$, reaches a maximum, then starts to broaden and decrease, ultimately splitting into two peaks with the stronger peak at $(\pi,\pi-\delta)$.  Figure~\ref{fig:6} plots the evolution of the peak (red solid line) and the onset of the split bands (blue line).  The blue squares are the data of Fig.~9 of the Main text, scaled by a factor $a=5$, showing that Eq.~\ref{eq:F1} is approximately satisfied.  The thin red lines represent constant-$\chi$ contours, labeled by the Stoner $U$ value.  Once again, as $U$ decreases, the unstable region converges onto the VHS doping.  We note that the onset of splitting (blue line in Fig.~\ref{fig:6}) is somewhat uncertain, as the peaks are quite broad and it is unclear just where they start to split.  

\subsection{Relation to spectrum-generating algebras}
Spectrum-generating algebras (SGAs)\cite{SGA1,SGA2} were developed in the context of nuclear physics to elucidate the underlying group structure of order parameters, taken as bilinears of creation and annihilation operators.  The primary concern of SGAs is not with interactions, but with the underlying topology of the order parameters, defined by the group structure of bilinears.  The Lie algebra contains chains of subalgebras, which define domains of attraction for the order parameters. Within each subalgebra, more conventional QCPs are driven by symmetry-breaking interactions which couple to individual elements of the subalgebra.   
SGAs are less concerned with the ordinary quantum critical points (QCPs) than with `supertransitions' across the separatrix between domains, brought about by external perturbations.  At this crossover, states of both subalgebras are degenerate, leading to divergent `order-parameter-DOSs' and possible emergent behavior. While phase transitions occur in SGAs at zero frequency, the domains and separatrices persist at all energies, leading to the concept of excited-state quantum critical points (ESQCPs).  One introduces a perturbation $\xi$ that switches the ground state from one subgroup to another, then plots the eigenvalue spectrum from one sector of the ground state (say, the states with angular momentum quantum number $ l = 0$) as $\xi$ is varied.  One thus finds a separatrix line in the $E-\xi$-space where the eigenvalues pile up, leading to a peak in the corresponding density-of-states.  As the eigenvalues cross this line, degenerate eigenvalues on one side split on the other side, signaling a change in the ground state.  As $E\rightarrow 0$, the separatrix extrapolates to the conventional QCP, but at finite $E$ it becomes an ESQCP.  

Attempts to extend the ideas of SGAs into condensed phases have run into difficulties.  A key difference between the SGAs in nuclear and condensed matter physics is that the former are designed with nearly-spherical nuclei in mind, and the associated groups are generally related to SO(N) and SU(N), whereas in solids, the pairs are first sorted by a $q$-vector, which is then superposed with the group of the star of each $q$.  

In Section SI.A we showed that the DFT-Lindhard suceptibility can be used to extend these ideas into the condensed matter domain.  Since the susceptibility $\chi_q$ can be identified with the order parameter $\sum_k<c^{\dagger}_{k+q}c_k>$, the fluctuation fingerprints bear a close resemblance to the domains of attraction in SGAs, with doping $x$ acting as a symmetry-changing parameter similar to $\xi$.  In this case, it is natural to identify the threshold value of the SDOS [corresponding to the largest susceptibility or the smallest inverse susceptibility] as the equivalent of the order-parameter DOS of SGA.  The Main text Figs.~14(c-e) show that the threshold SDOS diverges  at the Mott-Slater crossover. This divergence persists at finite $T$, suggestive of an ESQPT.  The emergent behavior we find at the transition is thus a form of deconfined QCPs (DQCPs), with a non-Landau-Ginzburg-Wilson transition between two types of competing order, where a new form of excitation emerges exactly at the DQCP.\cite{DQP1,DQP2}  We show in the Main Results Section (Beyond RVB) that the SDOS also diverges at the separatrix between $(\pi,\pi)$-order and the `ring' phase.

If we interpret the pseudogap as condensation into the coherent $(\pi,\pi)$-domain, then the low-temperature physics will be controlled by the SO(8) algebra of the VHS.\cite{SO8}  This algebra contains two SO(6) subalgebras of high interest to cuprate physics, one that combines Zhang's SO(5)\cite{SO5} [SDW plus d-wave superconductivity] with a flux phase operator, and a second with CDWs, spin-flux phase, and s-wave superconductors.  These two SO(6) vectors are coupled by a nematicity operator.  We see that this Lie algebra contains all of the low-$T$ phases found in the cuprate pseudogap.

In summary, we have constructed the condensed-matter analog of the spectrum-generating algebra, and shown that the characteristic frequency dependencies are echoed in the $T$-dependence, and hence experimentally accessible.  In particular, pseudogap physics in LSCO is dominated by the VHS domain of attraction, and the low-$T$ ordered phases are associated with the corresponding $SO(8)$-algebra.  

\section{Interpretation of Self-consistency}
\subsection{Relation to Entropy}
In Landau theory, fluctuations above the ground state are typically calculated in Gaussian approximation\cite{IvLy}, yielding an entropy in the paramagnetic phase
\begin{equation}
S=S_0(E)-\frac{1}{2T}\sum_{q,i}\chi^{-1}_{q}|\phi_{q,i}|^2,
\label{eq:F1}
\end{equation}
yielding a free energy
\begin{equation}
F=F_0+\frac{1}{2}\sum_{q,i}\chi^{-1}_{q}|\phi_{q,i}|^2,
\label{eq:F2}
\end{equation}
where $\phi_{q,i}$ ($i=x,y,z$) is the amplitude of the magnetic fluctuation.   Then if each degree of freedom of $F$ has an average energy of $T/2$, then $<|\phi_{q,i}|^2>=T\chi_{q}$, so the total fluctuation intensity is
\begin{equation}
\sum_{q,i}<|\phi_{q,i}|^2>=3T\sum_{q}\chi_{q}\propto\lambda.
\label{eq:F3}
\end{equation}
This clearly demonstrates that our calculation is a direct generalization of McMillan's phononic entropy\cite{25McMi} to the situation of electronic bosons.

\subsection{Relation to Bose-Einstein Condensation}

In the conventional picture of Bose-Einstein condensation (BEC), the total number of bosonic modes must satisfy
\begin{equation}
N=\int d\omega n_{BE}(\omega)D_B(\omega),
\label{eq:E1}
\end{equation}
where $n_{BE}$ is the BE distribution function and $D_B$ is the bosonic density of states.  As $T$ decreases, the bosonic chemical potential $\mu$ must adjust to maintain the equality.  However, the largest $\mu$ can be is zero, for which $n_{BE}\rightarrow n_P$, the Planck distribution, and the RHS of Eq.~\ref{eq:E1} takes on a fixed value.  For lower $T$, the equation can only be satisfied by putting a macroscopic number of bosons into the lowest energy mode.  

In the present situation, $D_B=\sum_q\chi''$, and $n_P$ can be approximated as $\sim T/\omega$, in which case Eq.~\ref{eq:E1} becomes
\begin{equation}
N\simeq T\sum_q\int d\omega \frac{\chi''(q,\omega)}{\omega} =\pi T\sum_q\chi'(q,\omega=0).
\label{eq:E2}
\end{equation}
Comparing this to Eq.~5 in the Main text, self-consistency has produced a number $N_{eff}=N\pi\lambda /A_0$ of `excitons' which can subsequently Bose condense.  Whereas for 3D BEC the macroscopic occupation of the lowest bosonic mode must be added by hand, in 2D the self-consistent condition automatically causes the occupation of that mode to diverge as $T\rightarrow 0$.

\subsection{Entropy and Dissipation}
As originally noted by McMillan\cite{25McMi}, anomalously large ratios of $2\Delta/k_BT_c$ for any phase transition are generically associated with a dominance of bosonic entropy at finite $T$ due to low-lying bosonic fluctuations.  But, from the fluctuation-dissipation theorem, the net fluctuations at a given frequency $\omega$ are equal to the dissipation $D(\omega)$ at that frequency,
\begin{eqnarray}
D(\omega)= \sum_q \chi''(q,\omega)(n_{BE}(\omega)+1/2)
\nonumber \\
\rightarrow T\sum_q \frac{\chi''(q,\omega)}{\omega}, 
\label{eq:E3}
\end{eqnarray}
so that the right-hand side of Eq.~\ref{eq:E2} is the total dissipation at temperature $T$.  In an ordered phase near $T=0$, the dominant fluctuations will be associated with Goldstone modes, but to show this in a calculation requires extending the present results to full self-consistency between the self-energy and vertex corrections as in a parquet calculation.

\section{Sample Mode Coupling Calculation}

Simple numerical integration of the Main text Eq.~14 cannot easily handle the logarithmic singularity near $X_-=U_c$.  Hence, we rewrite $N_-=N_1+N_2$, where $N_1$ has a simple analytical form and captures $N_-$ near the threshold, $N_2<<N_1$ for small $(X_--U_c)$.  Then the integral $I=\int N_-dX_-/(\delta +(X_- -U_c))$ is split into $I_1+I_2$, corresponding to the two $N_i$, and $I_1$ is evaluated analytically, while $I_2$ (which is nonsingular) is evaluated numerically.  For example, at higher $T$, $N_1$ varies approximately linearly with $X_-$ near the threshold, $N_1=N_a-N_b(X_--U_c)$ for $0<(X_--U_c)<X_c$.  In this case, 
 \begin{eqnarray}
I_1=(N_a-N_b\delta)ln(\frac{X_c}{\delta}+1)+N_bX_c. 
\label{eq:4b}
\end{eqnarray}
For small $\delta$ this reduces to the OZ form $I_1\sim N_aln(\frac{X_c}{\delta}),$ thereby describing Region I of the Main text Fig.~13(b).

For the small-$t'$ materials, the susceptibility on the plateau remains parabolic, but with a small curvature and a sharp cutoff.  This leads to an additional contribution to the SDOS of the form $N_-=N_p$ if $(X_--U_c)\le  X_p$. Then the integral of Eq.~\ref{eq:4b} contains an extra contribution
\begin{eqnarray}
I_p=N_p ln(\frac{X_p}{\delta}+1).
\label{eq:4c}
\end{eqnarray}
This term has two distinct limits.  At low $T$, $\delta<<X_p$, in which case $I_p=N_pln(X_p/\delta )$.  Combining this with $I_1$ leads to an OZ-like result of the same form but with a larger step height, $N_a\rightarrow N_a+N_p$, describing Region III of the Main text Fig.~13(b).
However, as $T$ increases, $\delta$ grows rapidly, leading to a possible reversal of the inequality.  
If $X_p<<\delta$, then $I_p\rightarrow N_pX_p/\delta$.  This is equivalent to assuming that the plateau has a flat top, in which case $N_-$ can be represented as a $\delta$-function, $N_-= N_pX_p\delta(X_--U_c)$, with $N_pX_p$ the excess height of $N$ at threshold, which is proportional to the area of the plateau at $T=0$.  From the Main text Eq.~14, $I_p\sim 1/\delta$ translates into $\xi_{th}\sim 1/T^{1/2}$, dotted line in the Main text Fig.~13(b), which approximately describes region II.  

\section{Relation to Other Calculations}

A comparison between our QPGW model and other `DFT+' calculations was given in Ref.~\onlinecite{AIP}; here we discuss this comparison specifically in regard to pseudogap physics.
A number of groups are working on MBPT-type extensions of DFT calculations.  However, DMFT and its cluster extensions are limited by a very low resolution in momentum $q$.  In particular, this means that even if one captures Mermin-Wagner physics, one will be typically limited to $\xi<7a$\cite{MaJar} [Fig. 50 of Ref.~\onlinecite{AIP}].  Furthermore, in averaging over large patches in momentum space, one loses the ability to resolve ordinary nesting instabilities, such as the ANN instability responsible for the cuprate CDWs [Fig. 49 of Ref.~\onlinecite{AIP}].  On the other hand, the fact that one can see the pseudogap, even when averaging over a quarter of the Brillouin zone is consistent with our finding that the pseudogap is spread over many separate modes in $q$.
Notably, whereas we find that the pseudogap is related to a peak in $\gamma$, recent cluster-DMFT calculations find that the pseudogap transition is associated with a line of peaks in the compressibility (Ref.~\onlinecite{AMST} and references therein). Since the Fermi liquid compressibility is proportional to the DOS, this line would also be expected to terminate at a VHS, consistent with our results. However, since the cluster-DMFT calculation assumed $t'=0$, the (paramagnetic) VHS would be at $x=0$, whereas the $T=0$ compressibility peak lies at a finite doping.  We suspect that it is related to the VHS of the bonding AF band. The cluster-DMFT calculation reveals an additional complication: within the superconducting state the compressibility diverges, leading to a regime of phase separation (see also Ref.~\onlinecite{GGVoz}).

Some DMFT extensions [dynamic vertex approximation\cite{DGA1,DGA3,DGA2}, dual fermion\cite{DF1,DF2}, and one-particle irreducible approaches\cite{OPI}] seek to calculate the two-particle vertex, and hence the momentum-dependent self energy, and are able to reproduce Mermin-Wagner physics and vertex corrections.  
In particular, a recent dynamical cluster approximation (DCA)\cite{flucdi} calculation sorted out the contribution of various channels of two-particle scattering to the self-energy (`fluctuation diagnostics'), and found that spin fluctuations are the origin of the pseudogap, while charge and pairing fluctuations play a marginal role, consistent with our results.  This has also been found experimentally\cite{Badoux}.

We further note that the parquet equations\cite{parq1,parq2} could be quite useful, if the recently discovered divergences\cite{flucdi} can be overcome. Mutual self-consistency between the self-energy and vertex corrections will be necessary to extend the current results to low temperatures.

After our paper was posted [arXiv:1505.04770], we became aware of several related calculations.  In particular, Ref.~\onlinecite{bottle} confirms the existence of an electronic bottleneck near the metal-insulator transition of the Hubbard model.  
An exact comparison is difficult, since their bottleneck arises in the midgap band of the dynamic mean-field theory doped Hubbard model, whereas in our quasiparticle-GW calculation this band arises due to nanoscale phase separation\cite{PeterNPS}.  Nevertheless, they found that the bottleneck occurs well inside the pseudogap phase, where the near-FS band starts to become coherent.  This is consistent with the transition in the Main text Fig.~13(b); note in particular that the bottleneck (red dot in the Main text Fig.~9) falls close to the coherent-incoherent transition (pink shaded region in Fig.~9 of the Main text).  
Ref.~\onlinecite{Pepin} also finds that strong bosonic fluctuations over a `quasidegenerate distribution of $2{\bf p_F}$ ordering wave vectors' drives the pseudogap, although they find it to be predominantly in the charge sector, forming a `resonant Peierls excitonic state'.